%% file: main.tex
\documentclass{article}
\usepackage[margin=1in]{geometry}
\usepackage{cite}
\usepackage[utf8]{inputenc}
\usepackage{amsmath,amssymb} %
\usepackage{xcolor}
\usepackage{float}
\usepackage{algorithm2e}
\usepackage{algorithmic}
\usepackage{subcaption}
\usepackage{csquotes}
\usepackage{url}
\usepackage{array}
\usepackage[thinlines]{easytable}
\usepackage{appendix}
\usepackage{ulem}
\usepackage{hyperref} 
\usepackage{booktabs}
\newcolumntype{L}{>{\centering\arraybackslash}m{4cm}}
\usepackage{lipsum}
\usepackage{graphicx}
\usepackage{caption}
\usepackage{breqn}
\usepackage{array}
\usepackage{geometry}
\graphicspath{{./Figures/}{../Figures}}
\RestyleAlgo{ruled}

\title{Topological Approach for Data Assimilation}

\author{
    \textbf{Max M.~Chumley}\\
    chumleym@msu.edu\\
    Michigan State University\\
    \and
    \textbf{Firas A.~Khasawneh$^*$}\\
    khasawn3@msu.edu\\
    Michigan State University\\
}

\date{\today}

\begin{document} 

\maketitle
*Address all correspondence to this author.
\input{./Sections/abstract}
\input{./Sections/intro}

\input{./Sections/theory}

\input{./Sections/methods}

\input{./Sections/results}

\input{./Sections/conclusion}

\section{Code Availability and Reproducibility}\label{sec:code}

We have made the TADA code publicly available in \texttt{teaspoon} \cite{Myers2020}, an open-source Python library for Topological Signal Processing (TSP). Documentation and example code are also provided to aid with reproducibility of the results.

\section{Acknowledgements}
This work is supported in part by Michigan State University and the National Science Foundation Research Traineeship Program (DGE-2152014) to Max Chumley.

\bibliographystyle{ieeetr}

\bibliography{Sections/bibliography}

\end{document}

%% file: Sections/abstract.tex
\section{Abstract}
Many dynamical systems are difficult or impossible to model using high fidelity physics based models. Consequently, researchers are relying more on data driven models to make predictions and forecasts. Based on limited training data, machine learning models often deviate from the true system states over time and need to be continually updated as new measurements are taken using data assimilation. Classical data assimilation algorithms typically require knowledge of the measurement noise statistics which may be unknown. In this paper, we introduce a new data assimilation algorithm with a foundation in topological data analysis. By leveraging the differentiability of functions of persistence, gradient descent optimization is used to minimize topological differences between measurements and forecast predictions by tuning data driven model coefficients without using noise information from the measurements. We describe the method and focus on its capabilities performance using the chaotic Lorenz 63 system as an example and we also show that the method works on a higher dimensional example with the Lorenz 96 system. 

%% file: Sections/intro.tex
\section{Introduction}
Physics-based dynamical system modeling is a very powerful and interpretable tool that can be used to make predictions and can provide relatively low-cost insight into system design and rapid prototyping. However, a model is only as good as the physics being used to define it and if the true system exhibits multi-scale behavior that requires extreme fidelity for simulation, the computational complexity outweighs the benefits of modeling the system. Many systems of interest to researchers are inherently multi-scale and high dimensional making them difficult or impossible to accurately predict using high fidelity physics based models. As a result, researchers are relying more heavily on data driven modeling techniques using machine learning to gain insight and make predictions for systems without the overhead computational cost \cite{montans2019data}. Many different data driven modeling techniques have been developed such as the AutoRegressive (AR), Moving Average (MA), and AutoRegressive Integrated Moving Average (ARIMA) models which assume a linear model form and learn coefficients from past training data \cite{zhang2003time,mouraud2017innovative}. A more modern approach to data driven modeling is rooted in deep learning and neural networks. While traditional Feed forward Neural Networks (FNN) are insufficient for time series forecasting due to the sequential ordering of points, Recurrent Neural Networks (RNN) are more suited for forecasting due to the dependence on previous states \cite{tanaka2019recent}. A form of RNN, the Echo State Network (ESN) has been used to construct data driven models from sparse measurements in \cite{yeo2019data}.  A modified version of the RNN that is used in forecasting is called Long Short Term Memory (LSTM) model which uses basic building blocks of an RNN to recall states from previous steps and has been shown to give significant improvements in forecast horizon compared to ARIMA \cite{siami2018comparison}. Another common forecasting approach is called Reservoir Computing (RC). RC works by mapping states into a high dimensional reservoir space and using linear regression to learn model coefficients as the features are mapped back into the original space \cite{tanaka2019recent}. RC based methods typically result in significant improvements to computation times while still accurately predicting future states of the system \cite{tanaka2019recent}. A special case of RC of interest for this paper is random feature map forecasting \cite{Gottwald_2021, gottwald2021combining}. This method is described in detail in Section~\ref{sec:raf_forecasting}. The quality of machine learning models is heavily dependent on the quality and quantity of training data. If the system changes states to a behavior that is drastically different form what was used to fit the coefficients, the corresponding model breaks down and the forecast will significantly deviate from the true system states. This issue is highlighted by any chaotic dynamical system where the forecast is accurate for a period of time after the training data and eventually deviates due to the finite model precision and training data. 

\begin{figure}[htbp]
  \centering 
  \includegraphics[width=0.4\textwidth]{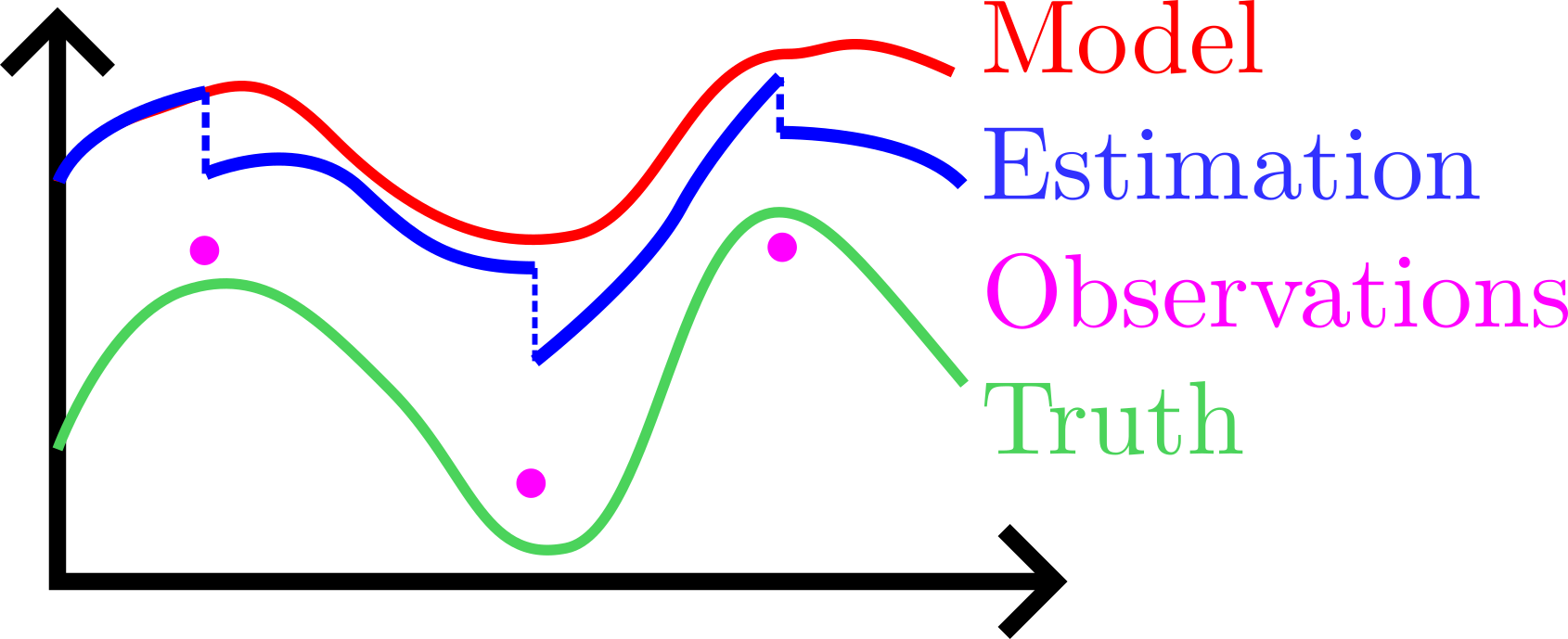}
   \caption{Data Assimilation Concept---Improving model results by considering observations to obtain an estimation closer to the ground truth.}
   \label{fig:da_schematic}
\end{figure}

To mitigate this problem, a concept called Data Assimilation (DA) is typically used. Data assimilation, or state estimation is a method for optimally combining observed data from multiple sources with model predictions to produce an improved prediction based on both \cite{ZHANG2015291, evensen2022data,dddas1, blasch2018dddas}. It has been successfully used across many fields such as weather forecasting, oceanography, predicting the movement of pollution and forecasting wild fires \cite{ZHANG2015291,li2017level, evensen2022data,cheng2022data}. 
In \cite{albarakati2024projected} a sliding window approach is taken using the Proper Orthogonal Decomposition to estimate the prominent structures in the data at the current window combined with DA techniques to obtain optimal predictions using few dimensions, but this approach can be sensitive to noise.   
A common implementation of data assimilation is the Kalman filter applied to dynamical systems where noisy system states measurements are optimally estimated using information from the noise statistics such as the measurement covariance matrix and a system forecast is generated by combining information from all available measurements \cite{evensen2022data, gottwald2021combining}. This process is demonstrated in Fig.~\ref{fig:da_schematic} where the observations move the model results closer to the ground truth to improve the predictions. 
In data assimilation, observed data streams and their uncertainties are used to update the model by solving for optimal weights with more importance given to sources with lower uncertainties. This is typically achieved by minimizing a cost function of the form  \cite{ZHANG2015291}
\begin{equation}\label{eq:da_cost}
    J(\vec{x}) = (\vec{x}-\vec{x}_b)^TB^{-1}(\vec{x}-\vec{x}_b)+(\vec{y}-h(\vec{x}))^TR^{-1}(\vec{y}-h(\vec{x})),
\end{equation}
where $\vec{x}_b\in\mathbb{R}^n$ is the vector of model prediction results, $B$ is the covariance matrix for the model, $\vec{y}\in\mathbb{R}^m$ is the vector of observations and $h$ is the operator that projects the input vector onto the space of observations, and $R$ is the covariance matrix for the measurements \cite{ZHANG2015291}. 
Thus finding $\vec{x}_a = \arg\min J(\vec{x})$ yields an optimal combination of the model predictions and measurements \cite{ZHANG2015291} where $x_a$ is referred to as the analysis solution. It was shown in \cite{ZHANG2015291} that $\vec{x}_a$ can be written as $\vec{x}_a=\vec{x}_b+K(\vec{y}-h(\vec{x}_b))$ for some weighting matrix $K$ in terms of the covariance matrices and the operator $h$. 
Random feature map forecasting and data assimilation have been combined using the ensemble Kalman filter in the Random Feature Maps and Data Assimilation (RAFDA) algorithm \cite{gottwald2021combining, Gottwald2016} which optimizes the model sequentially on the training data using ensemble sampling of points that follow the noise distribution. Other approaches also utilize the ensemble Kalman filter such as in \cite{jiang2022correcting} where the ensemble Kalman filter is used with dynamic mode decomposition to reconstruct dynamical systems from data. These methods assume a Gaussian white noise distribution with known statistics. Some filtering methods have been introduced for colored/correlated noise distributions in \cite{bryson1965linear, zare2021data} but they still rely on known noise statistics. In this paper, we assume that we do not have access to $R$ or $B$.

More recently, persistent homology from Topological Data Analysis (TDA) has been used to analyze complex data due to its compressive nature, vectorizability, and robustness to noise. Persistent homology was also shown to be robust to several noise distributions \cite{tempelman2020effects}, which can provide important advantages for data assimilation. 
However, the possibility of incorporating TDA into data assimilation was unlocked only recently with the definition of differential calculus on the space of persistence diagrams \cite{Leygonie2021,Carriere2020, Gameiro2016}. 
We leverage these advances in auto-differentiation of persistence to define a Topological Approach for Data Assimilation (TADA). This approach integrates random feature map forecasting with topological cost functions and persistence optimization to update the system model with incoming measurements. One of the key steps in TADA is generalizing the cost function in Eq.~\eqref{eq:da_cost} to use an arbitrary metric according to $J(\vec{x}) = d_b(\vec{x},\vec{x}_b)+d_y(\vec{y},h(\vec{x}))$, where $d_b$ is the model discrepancy and $d_y$ is the observation discrepancy. Using the Wasserstein distance as a metric to measure pairwise dissimilarity between the persistence diagrams of the measurement and model prediction enables the use of topology based cost functions \cite{li2018topological}. We show that our approach successfully produces system forecasts that are resilient to white, pink, and brownian noise. 

We organize the paper such that the relevant theoretical background on random feature map forecasting, PH, and persistence optimization is in Section~\ref{sec:theory}, the Topological Approach for Data Assimilation (TADA) algorithm is then presented in Section~\ref{sec:methods} with results in Section~\ref{sec:results} and conclusions in Section~\ref{sec:conclusion}.

%% file: Sections/theory.tex
\section{Theory}\label{sec:theory}
This section includes the necessary background on random feature map forecasting, TDA and persistence optimization.

\subsection{Random Feature Map Forecasting}\label{sec:raf_forecasting}

Random feature map forecasting is based on a machine learning approach that involves mapping the training data to high dimensional random features to learn a model in the new space \cite{NIPS2007_013a006f}. This approach has been used for time series forecasting in \cite{gottwald2021combining} where random features of the form $\phi(\mathbf{u})=\tanh{(\mathbf{W}_{in}\mathbf{u}+\mathbf{b}_{in})}$ where $\mathbf{W}_{in}\in\mathbb{R}^{D_r\times D}$ and $\mathbf{b}_{in}\in\mathbb{R}^{D_r}$ are the random weight matrix and bias vector for the features sampled from uniform distributions and are fixed for training \cite{gottwald2021combining}. 
$D$ is the system dimension and $D_r$ is the reservoir dimension or dimensionality of the random feature space. The vector $\mathbf{u}\in\mathbb{R}^D$ is the vector of system states and is assumed to come from a system of the form $\dot{\mathbf{u}}=F(\mathbf{u})$. Mapping the training data into the random feature space using $\phi$ allows for obtaining a surrogate model propagator map of the form $G=\mathbf{W}_{LR}\phi(\mathbf{u})$ where $\mathbf{W}_{LR}\in\mathbb{R}^{D\times D_r}$ optimally maps the random features back to the $D$ dimensional space to predict future states of the system. 
$\mathbf{W}_{LR}$ is obtained using ridge regression and the optimal solution is computed as $\boldsymbol{W}_{LR}=\boldsymbol{U}\boldsymbol{\Phi}^T(\boldsymbol{\Phi}\boldsymbol{\Phi}^T+\beta\boldsymbol{I})^{-1}$ where $\boldsymbol{U}\in\mathbb{R}^{D\times N}$ is a matrix of system states, $N$ is the number of training observations, $\boldsymbol{\Phi}\in\mathbb{R}^{D_r\times N}$ is the matrix of random features $I$ is the $D_r\times D_r$ identity matrix and $\beta$ is the regularization parameter \cite{Chatfield_2000}. 

\subsection{Topological Data Analysis}
This section reviews the needed concepts from topological data analysis: persistent homology (Section~\ref{sec:persistence_background}) and persistence optimization (Section~\ref{sec:persistence_opt}). Topological data analysis (TDA) quantifies structure in data. 
We review the basics of persistent homology for point clouds in $\mathbb{R}^n$, but more specifics can be found in \cite{Hatcher2002,Kaczynski2004,Ghrist2008,Carlsson2009,Edelsbrunner2010,Mischaikow2013,oudot2017persistence,Munch2017}. While the theory is presented in terms of a general point cloud, it is helpful to think of the points being states of an $n$-dimensional dynamical system.

\subsubsection{Persistent Homology}\label{sec:persistence_background}

\begin{figure}[htbp]
    \centering
    \includegraphics[width=0.95\textwidth]{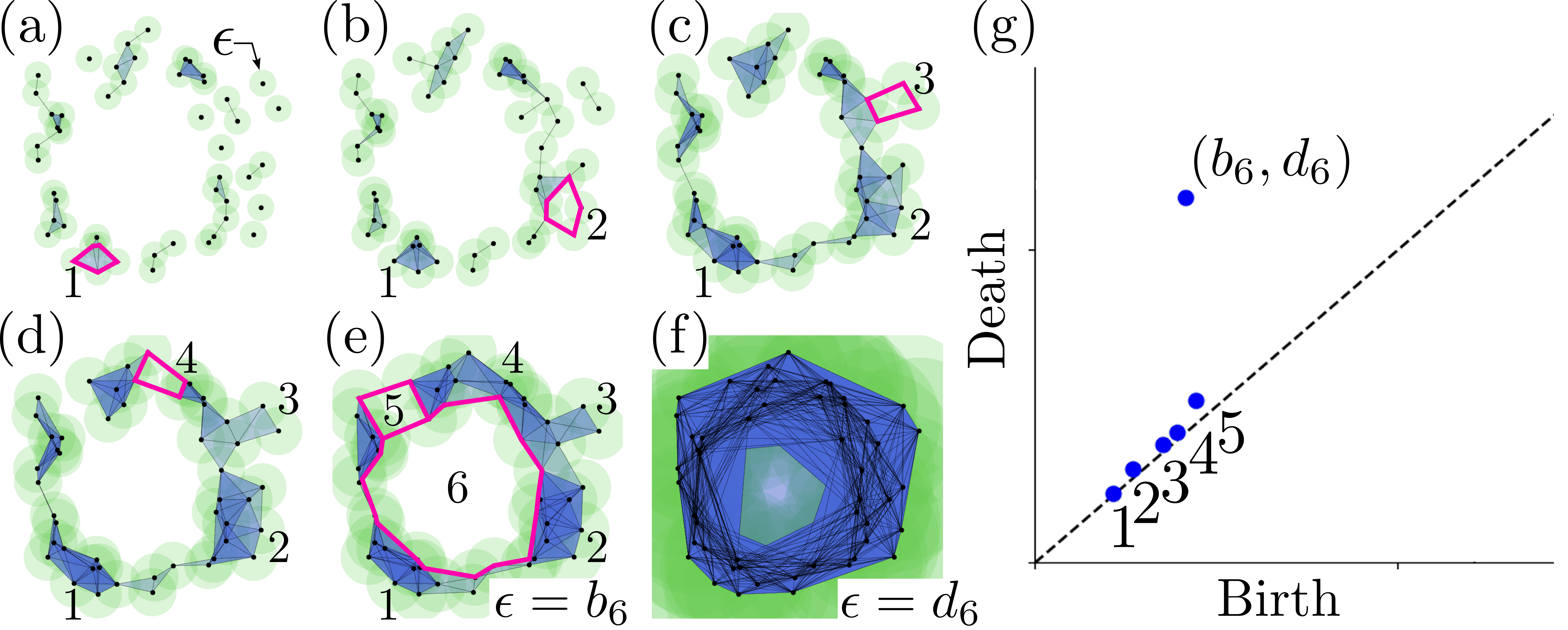}
    \caption{Persistence for point clouds. Each snapshot in (a)--(e) shows the rips complex for increasing values of a disc of radius $\epsilon$. One prominent loop is formed (or born) at $\epsilon=b_6$ in (e), and fills in (or dies) in (f) when $\epsilon=d_6$. The other 5 loops are small and a result of noise so they are born and die at nearly equivalent values of $\epsilon$. These loops are represented in the 1D persistence diagram (g) as (birth,death) pairs. Non-prominent loops form and die quickly as shown by the points near the diagonal.}
    \label{fig:Rich_Complex}
\end{figure}

Homology groups can be used to quantify structure or shape in different dimensions on a simplicial complex $K$. This is done using homology, $H_p(K)$, which is a vector space computed from the complex, where $p$ is the dimension of structures measured.
For example, in dimension $0$, the rank of the $0$ dimensional homology group $H_0(K)$ is the number of connected components. 
The rank of the $1$-dimensional homology group $H_1(K)$ is the number of loops or holes, while the rank of $H_2(K)$ is the number of voids, and so on. For example, consider Fig.~\ref{fig:Rich_Complex}(e) where we see that the simplicial complex contains two holes meaning that the rank of the 1D homology at this particular value of the connectivity parameter is 2.

We are interested in studying the structure of a changing simplicial complex (a generalization of a graph) by measuring its changing homology. This process is called persistent homology. 
In general, we assume we have a real valued function on the simplices of $K$ whose values are used to generate the simplicial complex. For this paper we focus on a specific simplicial complex called the Vietoris-Rips or simply Rips complex (VR) where the function on the point cloud $\{ x_1,\cdots,x_N\} \subseteq \mathbb{R}^d$ becomes the euclidean distance between points. The basic idea is that the Rips complex with parameter $\epsilon$ is a higher dimensional analogue of the proximity graph, where two vertices are connected with an edge if the distance between the relevant points in the point cloud are at most distance $\epsilon$. 
This process forms a nested sequence or filtration of simplicial complexes $K_0\subseteq K_1\subseteq ... \subseteq K_n$ which induces a sequence of inclusion maps on the homology $H_p(K_1) \to H_p(K_2)\to \cdots \to H_p(K_n)$. An example can be seen in the series of simplicial complexes in Fig.~\ref{fig:Rich_Complex}.

The appearance and disappearance of holes in the filtration is encoded in this sequence. 
This information is then represented in a persistence diagram, where the large loop appears at $\epsilon=b_6$ and fills in at $\epsilon=d_6$. This information is represented as a point in  $\mathbb{R}^2$ at $(b_6,d_6)$ in Fig.~\ref{fig:Rich_Complex}(f). The other 5 loops in this particular point cloud are all born and die almost immediately so those persistence pairs show up near the diagonal. 
The collection of the points in the persistence diagram give a summary of the topological features that persist over the defined filtration. 
Points far from the diagonal represent structures that persist for a long time, and thus are often considered to be prominent features. Conversely, points close to the diagonal are often attributed to noise in the data and it is clear that loops 1--5 are due to the noise in Fig.~\ref{fig:Rich_Complex}.

\subsubsection{Persistence Optimization}\label{sec:persistence_opt}
An emerging subfield of topological data analysis deals with optimization of persistence based functions by exploiting the differentiability of persistence diagrams. Persistence diagrams are commonly represented by many different scalar features used for machine learning such as the total persistence \cite{Carriere2020},
$E(D) = \sum_{i=1}^p |d_i-b_i|$,
which gives a measure of how far the persistence pairs are from the diagonal. In other words, this feature gives the sum of the persistence lifetimes $\ell_i=d_i-b_i$. These scalar representations are referred to as \textit{functions of persistence} \cite{Carriere2020}. Other examples of functions of persistence include maximum persistence, $E(D) = \max_i|d_i-b_i|$, and persistent entropy $E(D)=-\sum_i p_i\log_2(p_i)$ where $p_i=\frac{\ell_i}{\sum_i \ell_i}$ \cite{chintakunta2015entropy} which gives a measure of order of the persistence diagram. 
Features such as the Wasserstein or bottleneck distance can also be used for measuring dissimilarities between two PDs \cite{Carriere2020, schrader2023topological}. In \cite{Carriere2020}, it is specified that in order to have differentiability of the persistence map, the function of persistence must be locally Lipschitz and definable in an o-minimal structure or in other words definable using finitely many unions of points and intervals. An example of a set that fails this criteria is the cantor set because it requires infinitely many operations to determine if a point is in the set.

\begin{figure}
    \centering
    \includegraphics[width=0.9\textwidth]{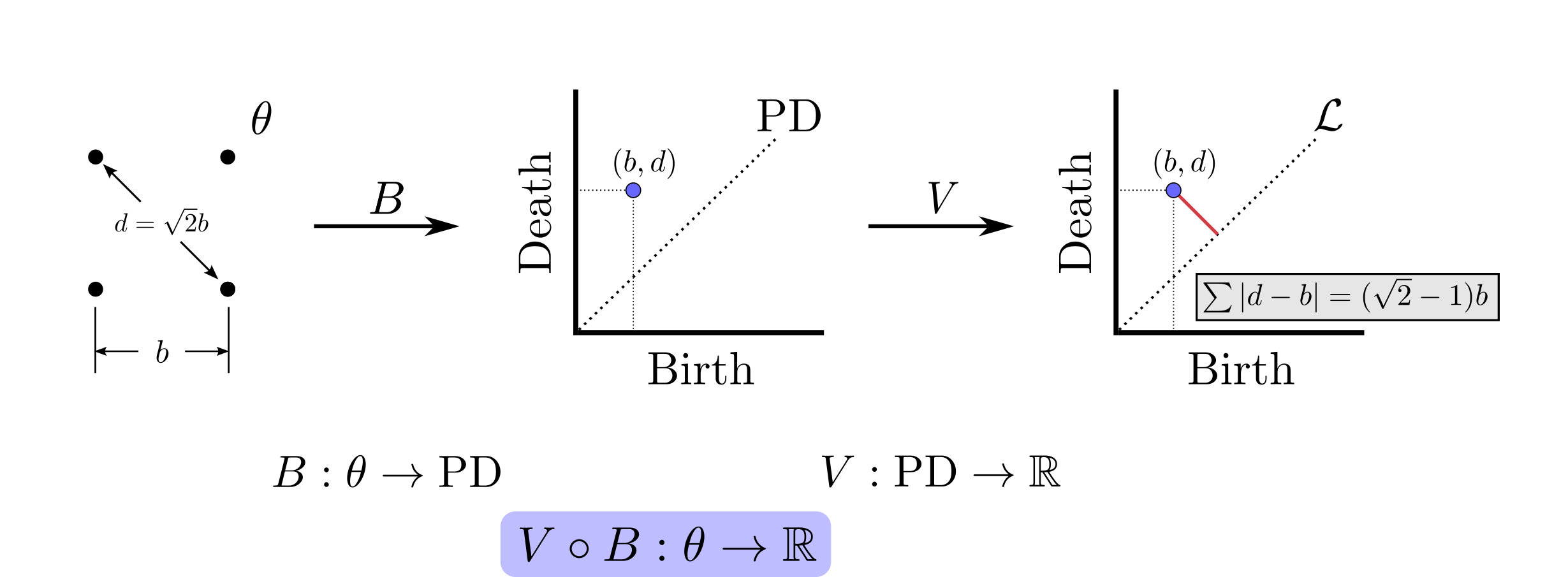}
    \caption{Mapping a point cloud $\theta$ to a real values persistence feature using the map composition $V\circ B$.}
    \label{fig:pers_map}
\end{figure}

Generally, a function of persistence is evaluated through the map composition, 
\begin{equation}
    \mathcal{C}:\mathcal{M} \xrightarrow[\hspace{1.1cm}]{B} \text{PD} \xrightarrow[\hspace{1.1cm}]{V} \mathbb{R},
\end{equation}
where the input space $\mathcal{M}$ can be a point cloud or image that is mapped to a persistence diagram using the filtration $B$ \cite{Leygonie2021}. An example of this mapping is shown in Fig.~\ref{fig:pers_map} where the square point cloud is mapped to a persistence diagram using the map $B$ with VR filter function and the persistence diagram $PD$ is mapped to the total persistence feature using the map $V$. The composition of these maps ($V\circ B$) allows for directly mapping the point cloud to a persistence features. 
Reference~\cite{Carriere2020} outlines the optimization of persistence-based functions especially via stochastic subgradient descent algorithms for simplicial and cubical complexes with explicit conditions that ensure convergence. A function of persistence is defined as a map from the space of persistence diagrams associated to a filtration of a simplicial complex to the real numbers such that it is invariant to permutations of the points of the persistence diagram.

\begin{figure}
    \centering
    \includegraphics[width=0.9\textwidth]{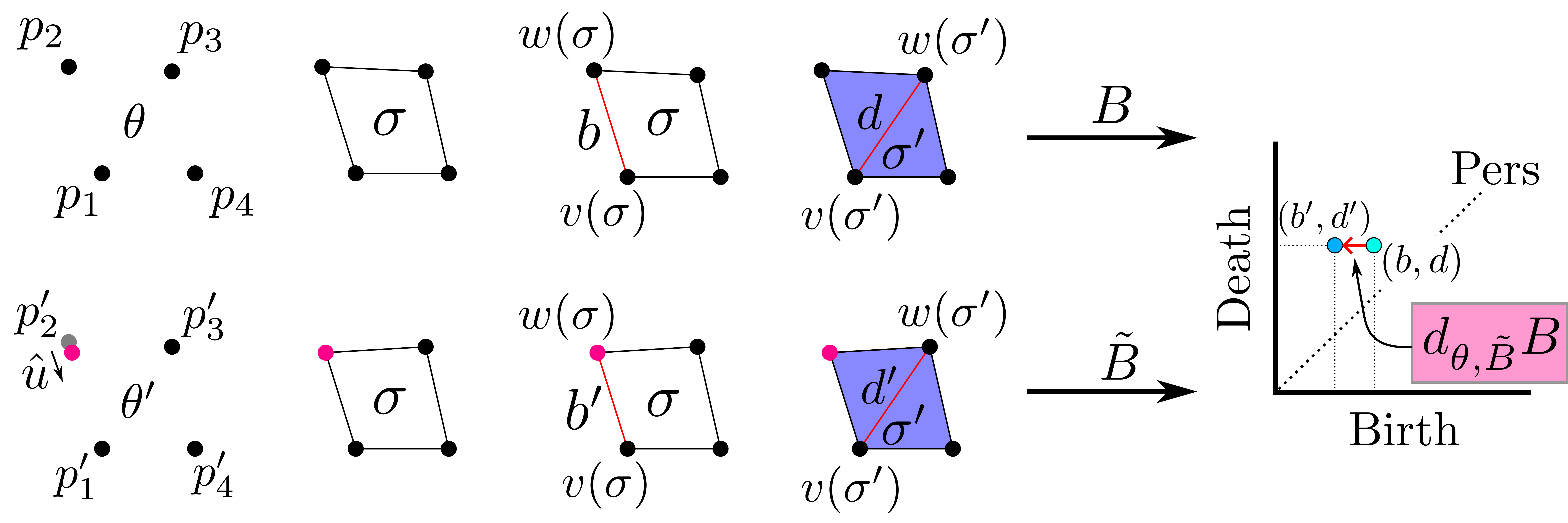}
    \caption{Persistence diagram differentiation process. The top row shows the process of tracking the birth and death of the loop from the original point cloud along with using the map $B$ to obtain its persistence diagram. The bottom row performs the same process on a perturbed point cloud and demonstrates how the change in the persistence pair forms the derivative $d_{\theta,\tilde{B}}B$.}
    \label{fig:pd_der1D}
\end{figure}

The PD is represented as a $\mathbb{R}$ number by way of the chosen function of persistence $V$. $\mathcal{C}$ has enabled differentiability and gradient descent optimization of its members using the chain rule on $V\circ B$ to obtain desired characteristics of $\mathcal{M}$ \cite{Carriere2020,Leygonie2021, Gameiro2016}. 
$B$ is differentiated by considering a local perturbation or \textit{lift} of the input space $\mathcal{M}$, $\tilde{B}$. The space of possible perturbations is then mapped onto the PD, and for a particular perturbation of $\mathcal{M}$, the directions of change of the persistence pairs form the derivative of $B$ with respect to $\tilde{B}$ \cite{Leygonie2021}. 
This process is pictorially represented using a simple point cloud in $\mathbb{R}^2$ consisting of a single loop in Fig.~\ref{fig:pd_der1D}. The top row from left to right shows the original point cloud along with the simplicial complex where the loop is born $\sigma$, and where it dies $\sigma '$. The corresponding \emph{attaching edges} where these events occur are labeled as $b$ and $d$ with vertices $w(\cdot)$ and $v(\cdot)$. The map $B$ is used to map the point cloud to the persistence diagram. The bottom row of Fig.~\ref{fig:pd_der1D} demonstrates the same process as the top row but on a perturbed point cloud $\theta '$ where $p_2\to p_2'$ along $\hat{u}$. The map $\tilde{B}$ represents the persistence map for the perturbed point cloud and the resulting change in the persistence pair forms the derivative of the persistence map $B$ with respect to $\theta$ and $\tilde{B}$. Notationally, this is represented as $d_{\theta,\tilde{B}}B$ \cite{Leygonie2021}.

\begin{figure}[htbp]
    \centering
    \includegraphics[width=0.35\textwidth]{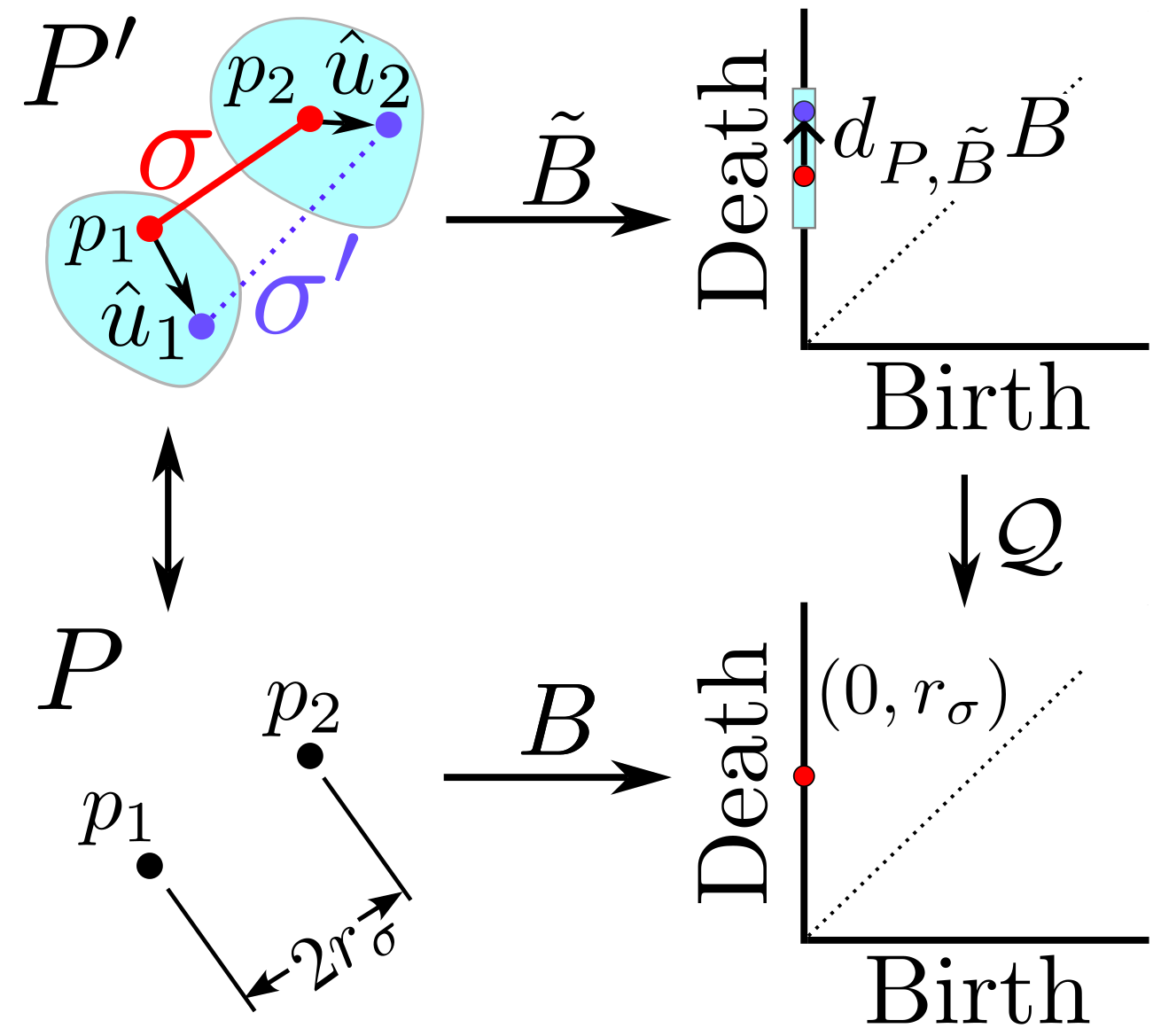}
    \caption{Persistence differentiation for point clouds. The point cloud $P$ is perturbed to $P'$ and the 0D persistence diagram is differentiated with respect to this perturbation.}
    \label{fig:pd_diff}
\end{figure}

This process is illustrated more generally and for 0D persistence in the example shown in Fig.~\ref{fig:pd_diff}. In this diagram, the space of infinitesimal perturbations of the point cloud $P$ is shown in blue and this higher dimensional space is mapped onto a persistence diagram where the quotient of the space collapses to the original persistence pair. For the particular perturbation shown we see that the edge length is increasing so the derivative of $B$ using the VR filtration with respect to the perturbation $P'$, the corresponding persistence map $\tilde{B}$ is a vector in the vertical direction. For higher dimensional simplices or PDs such as in Fig.~\ref{fig:pd_der1D}, the process is the same, however, we consider the rate of change of the attaching edge of the simplex or the edge whose inclusion results in the birth of the simplex \cite{Carriere2020,Leygonie2021}. Attaching edges are the output of the corresponding filter function chosen. 
For example, if the VR filtration is used, the filter function for a simplex $\sigma$ is defined to be $F(P)(\sigma)=\max_{i,j\in\sigma}{||p_i-p_j||_2}$ or the maximal distance between any two vertices in the simplex \cite{Leygonie2021} where $||\cdot||_2$ is the $l_2$ norm. Before the connectivity parameter reaches $F(P)(\sigma)$, $\sigma$ remains unborn in the filtration. In this case, the map $B$ corresponds to the composition of the persistence map $\text{Dgm}_p$ and the filter function $F$ \cite{Leygonie2021}. Conditions of differentiability must be considered for the input space being studied. 
If the input is a point cloud it must be in \emph{general position} \cite{Leygonie2021, Gameiro2016} (i.e., no two points in the cloud coincide or are equidistant). Nonetheless, if the general position condition fails then the derivative likely still exists for the specified perturbation. The issue is also mitigated numerically by cpu floating point precision and the constraints are highly unlikely to be violated with real data \cite{Gameiro2016}. If either condition is violated, small artificial noise can also be introduced to guarantee the points are in general position and a unique perturbation exists.

For point cloud input data, the derivative of a persistence diagram is computed by labeling the vertices of attaching edges for the birth $\sigma$ and death $\sigma'$ of a simplex as $v(\sigma)$, $w(\sigma)$ and $v(\sigma')$, $w(\sigma')$ respectively for each attaching edge that results in the birth or death of a homology class. An arbitrary perturbation $P'$ of the point cloud $P$ is then considered. The attaching edge vertices are tracked in the process allowing for each persistence pair to measure the direction of change and construct the derivative of the persistence diagram with respect to that perturbation. 
The derivative is represented by a list of vectors (one for each persistence pair) that describe the variation in the persistence pairs with respect to a given perturbation $P'$. A unit direction vector $\hat{u}$ is used to store the perturbation directions for each point. The derivative is then computed via an inner product of the vector $P_{i,j}=\frac{p_i-p_j}{||p_i-p_j||_2}$ which describes the direction of change in length of the attaching edge $(i,j)$ and the perturbation vector $\hat{u}$. Mathematically using the VR filtration, the derivative takes the form,
\begin{equation}\label{eq:pd_der}
    d_{P,\tilde{B}}B(\hat{u})=\left[\left(P^T_{v(\sigma),w(\sigma)}\hat{u},~P^T_{v(\sigma '),w(\sigma ')}\hat{u}  \right)_{i=1}^m\right],
\end{equation}
where $d_{P,\tilde{B}}B$ is the derivative of the persistence map $B$ with respect to the perturbation persistence map $\tilde{B}$ evaluated at the perturbation $\hat{u}$ \cite{Leygonie2021}. Note that the form Eq.~\eqref{eq:pd_der} has been represented for $m$ finite persistence pairs generalizations are presented in \cite{Leygonie2021} from parameterization by Rips filtration to present a formal framework for differentiation of persistence diagrams using maps between smooth manifolds $\mathcal{M}$ and $\mathcal{N}$ through space of persistence diagrams with a general filter function in \cite{Leygonie2021}. This framework also includes generalizations to infinite persistence pairs, however, for this work we are mainly interested in finite persistence pairs using the VR filter function.

One of the primary applications of this optimization comes from \cite{Carriere2020} where a \texttt{TensorFlow} pipeline was developed using the \texttt{Gudhi} TDA library in python to optimize the positions of points in a point cloud with gradient descent according to a predefined loss function. The loss function in \cite{Carriere2020} was defined to maximize the total persistence or in other words expand the size of the loops in the 1D persistence diagram. A term was also added to the cost function to regularize by restricting the points to a square region of space. More loss functions can also be defined in terms of persistent entropy to promote fewer loops in the point cloud and using the Wasserstein distance to achieve a desired persistence diagram. 
The work in \cite{Gameiro2016} outlines processes for carrying out optimization using persistence based functions in the specific case of Vietoris-Rips complexes defined on point clouds. Particularly, these methods allow for the user to supply a start and end persistence diagram along with the starting point cloud. Gradient descent is then used to optimally transform the original point cloud into a new point cloud that has the desired homology.

%% file: Sections/methods.tex
\section{Topological Approach for Data Assimilation (TADA)} \label{sec:methods}

To efficiently combine data without knowing the noise distribution of the measurements, we integrate the random feature map forecasting method with a new data assimilation scheme driven by TDA to optimally combine the learned model from each signal taking into account data uncertainties in the form of topological differences. We define the cost function in terms of common discrepancy measures between persistence diagrams such as the Wasserstein distance similar to \cite{li2018topological}. Persistence optimization is then used to compute a new, optimal forecast model for all of the input time series signals. We call our method Topological Approach for Data Assimilation or TADA. 

\begin{figure}[htbp]
  \centering 
    \includegraphics[width=0.9\textwidth]{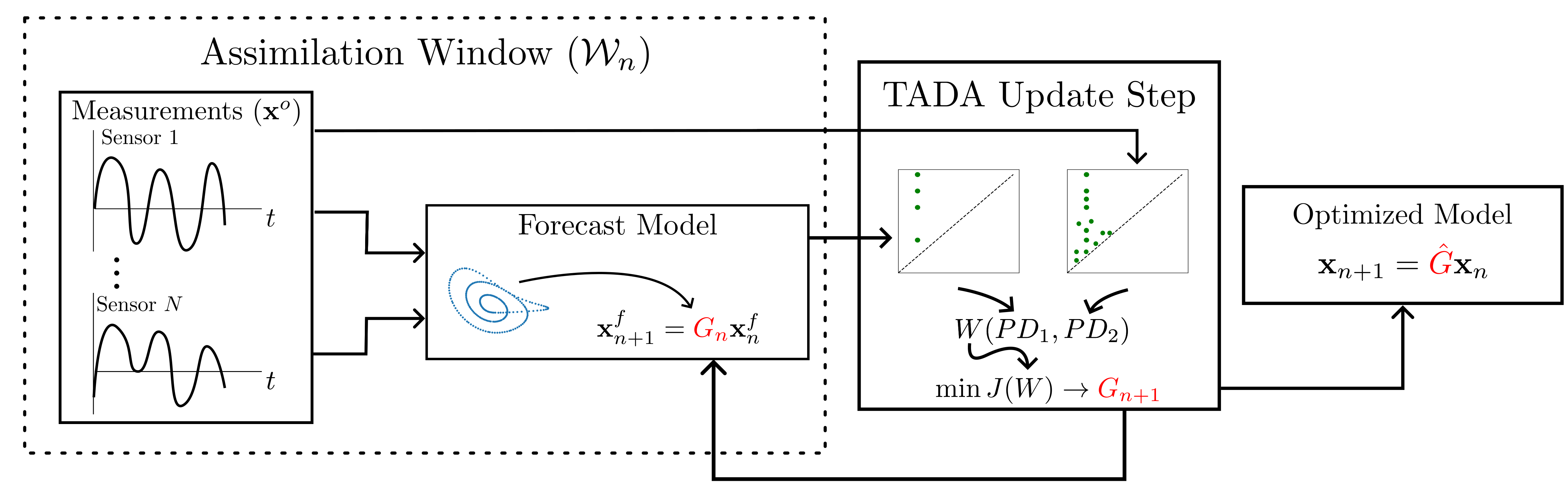}
    \caption{Assimilation window update diagram showing the updated model giving an improved forecast.}
    \label{fig:td_learning}
\end{figure}

Given $N$ sensor observations with additive noise, we generate a random feature map model $G_0=W_{LR}$. We use this model to forecast the system response $\mathcal{W}$ points into the future and we collect state measurements over the same forecast window. This gives us for the $n$th data assimilation window $\mathcal{W}_n$ two point clouds corresponding to the forecast states and their measured counterpart. We compute the persistence diagrams for these two point clouds and use them within persistence optimization to update the weights of the forecast model so that it better fits the observed response. 
To update the model $G_0$ using the $n$th assimilation window, the model coefficients are varied such that the differences between the persistence diagrams of the predicted and measured states are minimized over $\mathcal{W}_n$. This is achieved using a cost function $J_1(W,\hat{W}):~\mathbb{R}^4\to\mathbb{R}$ where $W=(W_0,W_1)^T$ is the vector with components consisting of the 0D and 1D Wasserstein distances between the model and measurement persistence diagrams, respectively. Similarly, $\hat{W}$ is the vector containing Wasserstein distances between the empty persistence diagram and the model error persistence diagram or the persistence diagram of the point cloud generated by taking the differences between the model predictions and measurements. We found that both $W$ and $\hat{W}$ were important because if only $W$ is used many solutions exist to the optimization problem with similar but time-shifted shape as the measurements leading to undesirable temporal shifts in the predictions. Note that the Wasserstein distance between two persistence diagrams is computed using,
\begin{equation}
    W_p(PD_1, PD_2) =  \inf_\pi \left( \frac{1}{n} \sum_{i=1}^n \|X_i - Y_{\pi(i)}\|^p \right)^{\frac{1}{p}} ,
\end{equation}
where $PD_1$ and $PD_2$ are the persistence diagrams in the desired dimension, $X_i$ are the persistence pairs of $PD_1$ and $Y_{\pi (i)}$ are persistence pairs of $PD_2$ optimally matched to $PD_1$ pairs to minimize the total distance (using a $\ell_2$-norm metric) between the points with optimal transport. For this work we use $p=1$ for simplicity. We take $J_1$ to be the sum of the 0D and 1D Wasserstein distances for $W$ and $\hat{W}$. Note that the model persistence diagrams are inherently functions of the model $G_n$ so minimizing $J_1$ will yield an updated forecast model $G_{n+1}$ to get the analysis solution for the next assimilation window. The next assimilation window is obtained by taking the next $\mathcal{W}$ predictions and measurements shifted by a stride length $\mathcal{S}$. To minimize topological changes between windows, we take $\mathcal{S}=1$ for this paper. We apply this pipeline over many assimilation windows, and obtain an optimized estimator $\hat{G}$. This approach is pictorially represented in Fig.~\ref{fig:td_learning}. Note that the size of the very first window starts with just two points and as new measurements are obtained it increases to the specified window size and slides across the signal after that point.

In order to ensure that the model does not deviate from the training data, we include another term to the cost function to constrain the problem to the training set using persistence optimization. The full cost function becomes $J=J_1(W,\hat{W}) + J_2(W,\hat{W})$, where $J_2$ minimizes topological differences between randomly sampled model and measurement point clouds of fixed size on the training set. As the window slides along the signals measurements that exit the window are then added to the set of points that can be randomly sampled for $J_2$. Together, these two cost function terms allow the model to converge to the original model by minimizing errors on the training set, but also allows the model to learn from incoming data with the sliding window approach. Unless otherwise specified, for all results in this paper we use the full cost function and randomly sample 100 points from the training set for $J_2$, but this number can be varied depending on the number of training points.

\begin{figure}[htbp]
  \centering 
    \includegraphics[width=0.8\textwidth]{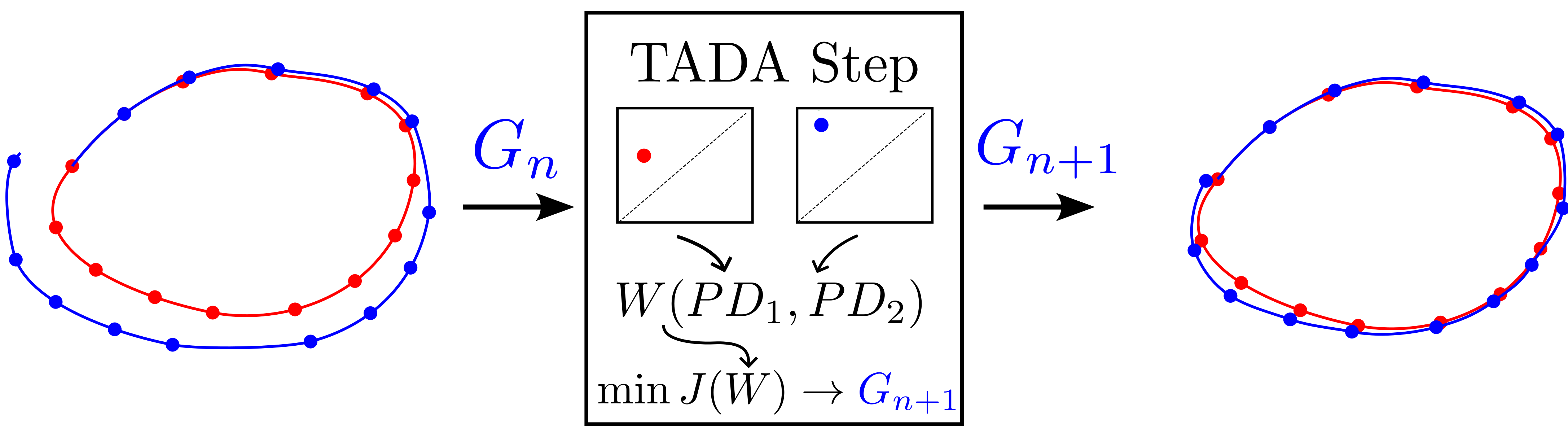}
    \caption{Assimilation window update diagram showing the updated model giving an improved forecast.}
    \label{fig:window_update}
\end{figure}

This approach does not contain any restriction on the noise model in the measurements, and when combined with forecasting methods that are also independent of the noise characteristics in the signal it enables a noise agnostic data assimilation. Figure~\ref{fig:window_update} shows a snapshot of an assimilation window where the forecast (blue) deviates from the measurements (red). Persistence optimization is used in the second box to minimize the Wasserstein distance between the persistence diagrams giving a forecast model that is much closer to the measurement. The process then repeats to update the forecast weights for the next window.

%% file: Sections/results.tex
\section{Results}\label{sec:results}

\subsection{Lorenz System}
In this paper, we mainly focus on testing TADA using the chaotic Lorenz 63 dynamical system defined as, $\dot{x} = \sigma (y - x)$, $\dot{y} = x (\rho - z) - y$, $\dot{z} = xy - \beta z$ where $\sigma$, $\rho$, $\beta$ were set to $28$, $10$ and $8/3$ respectively for chaotic dynamics. Simulations were conducted using randomly selected initial conditions by sampling from the standard normal distribution. Simulations were sampled at a frequency of 50 Hz for 500 seconds and the last 6000 points were taken to remove transient behavior. Noise was added to the simulation signals using the form from \cite{gottwald2021combining} where the observational error covariance matrix is defined as $\Gamma$ and the noise is added as $\Gamma^{1/2}\boldsymbol{\eta}$ where $\boldsymbol{\eta}\in\mathbb{R}^D$ is sampled from the noise distribution. In \cite{gottwald2021combining}, the authors take $\Gamma=\eta I$, however this effectively applies different noise levels to each system state due to the differing amplitudes. Therefore, we chose to apply different noise amplitudes $\eta_i$ for each state based on a specified Signal-to-Noise Ratio (SNR) in decibels (dB) using,
\begin{equation}
    \eta_i = A_{\text{signal}}^i 10^{-\text{SNR}_{\text{dB}}/20},
\end{equation}
where $A_{\text{signal}}^i$ is the RMS amplitude of signal $i$ and $\text{SNR}_{\text{dB}}$ is the signal-to-noise ratio in decibels. This makes $\Gamma$ a diagonal matrix with different noise amplitudes to apply the same noise level to each state.

\subsection{Random Feature Map Parameters}

Data driven models were generated for each simulation of the Lorenz system using the random feature map method from Section~\ref{sec:raf_forecasting}. For consistency, we chose to conduct all tests using a reservoir dimension of 300 ($D_R=300$). Random weights were sampled from the uniform distribution $\mathcal{U}(-0.005,0.005)$ and the bias vector entries were sampled from $\mathcal{U}(-4.0,4.0)$. The regularization parameter, $\beta$, was set to $4\times 10^{-5}$. Accuracy of the random feature map method is heavily dependent on the parameters chosen for the random features and regularization \cite{Gottwald_2021}. In this work, we use the same parameters from \cite{Gottwald_2021} specific to the chaotic Lorenz system, but choosing these parameters for an arbitrary system is highly nontrivial. In all simulations, we used 4000 points for training the random feature map model.

\subsection{Forecast Time}
To quantify the accuracy of the forecasts and assimilation updates, we use the relative forecast error,

\begin{equation}\label{eq:fc_err}
    \mathcal{E}(t_n)=\frac{||\boldsymbol{u}_{\text{valid}}(t_n)-\boldsymbol{u}_n(t_n)||^2}{||\boldsymbol{u}_{\text{valid}}(t_n)||^2},
\end{equation}
where $\boldsymbol{u}_{\text{valid}}(t_n)$ is the validation set or measurement
data and $\boldsymbol{u}_n(t_n)$ is the analysis or optimal estimation. The
forecast horizon is computed based on a threshold of the forecast error. The forecast time $\tau_f$ is the maximum time such that 
$\mathcal{E}(\tau_f)<\theta$. In
\cite{gottwald2021combining} $\theta$ was taken to be 0.05 and the 2-norm was
used for all computations. It is standard practice to quantify DA results in terms of Lyapunov times where the forecast time is scaled by the maximum Lyapunov exponent of the system $\lambda$. In this paper, all results are from the Lorenz system where the maximum Lyapunov exponent was taken to be $\lambda=0.91$ \cite{gottwald2021combining}.

\subsection{TADA Forecast}

\begin{figure}[htbp]
    \centering
    \includegraphics[width=0.45\textwidth]{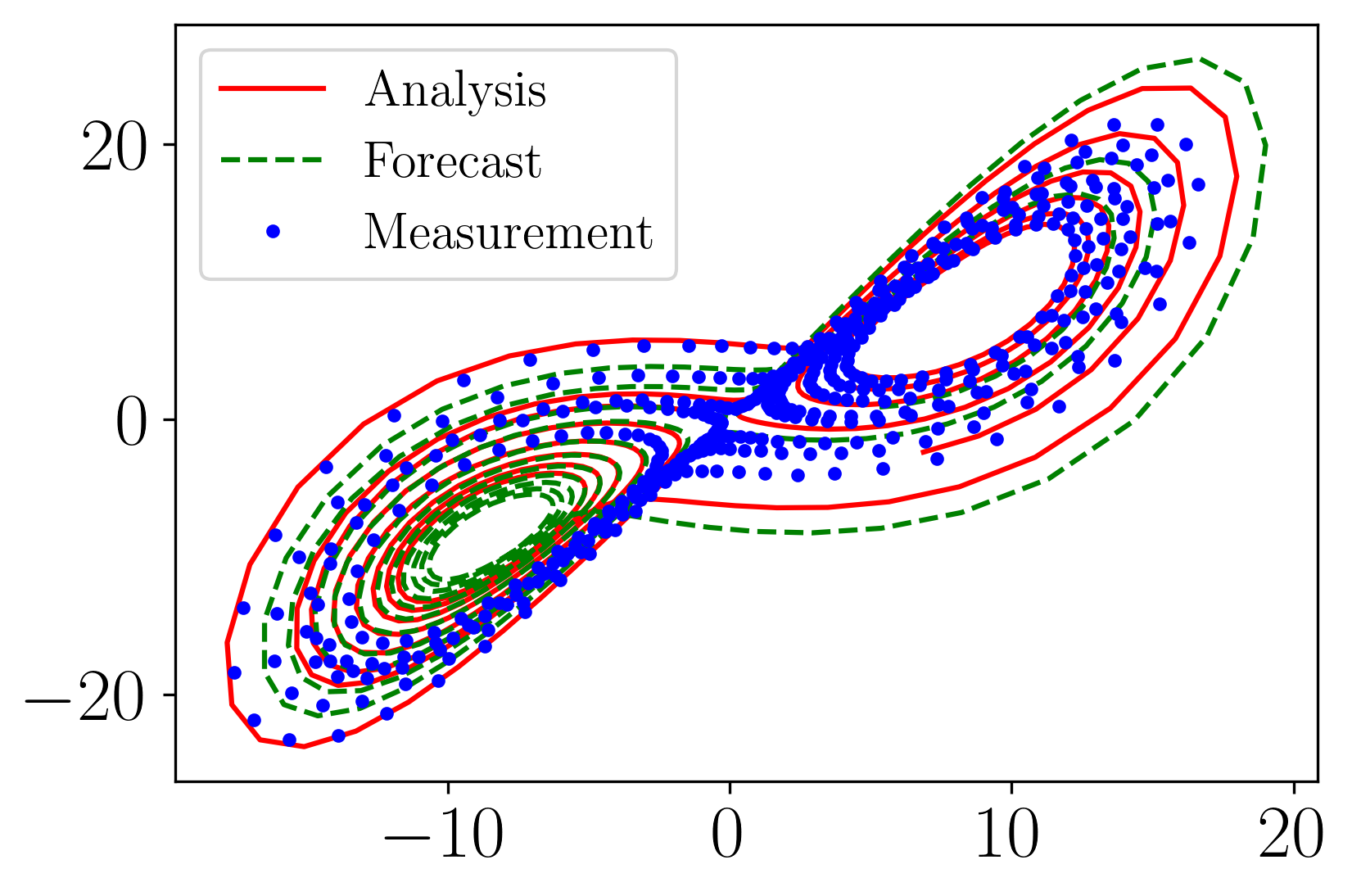}
    \caption{Two dimensional state space projection of noisy Lorenz system measurements (blue points) with the initial forecast (green dashed) and the TADA forecast (red solid). The initial forecast time was 0.89 and the TADA forecast time increased to 4.08 Lyapunov time units.}
    \label{fig:state_space_DA}
\end{figure}

\begin{figure}[H]
    \centering
    \includegraphics[width=0.95\textwidth]{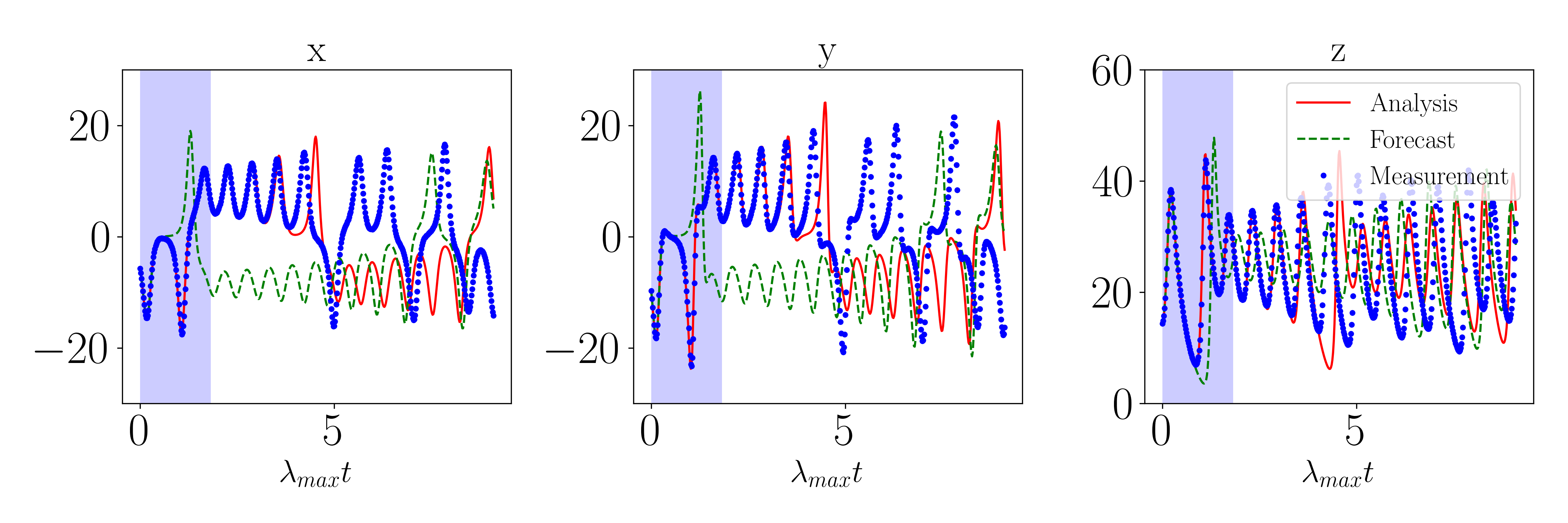}
    \caption{Time domain plots of noisy Lorenz system measurements (blue points) with the initial forecast (green dashed) and the TADA forecast (red solid). The initial forecast time was 0.89 and the TADA forecast time increased to 4.08 Lyapunov time units. The blue regions indicate the measurements that were used to improve the forecast. }
    \label{fig:time_domain_DA}
\end{figure}

The TADA algorithm was first applied to a single Lorenz system trajectory to demonstrate a successful DA update. We used a window size of 50 points with 50 optimization epochs so 100 new measurements are used for data assimilation. In Fig.~\ref{fig:state_space_DA} we show a two dimensional projection of the state space trajectory just after the training data. The blue points are the measurements, the dashed green curve is the initial forecast obtained using random feature maps and the solid red curve is the analysis solution after 50 data assimilation updates. We see that the analysis solution appears to follow the measurements better generally. For the trajectory in Fig.~\ref{fig:state_space_DA}, the corresponding time domain results are shown in Fig.~\ref{fig:time_domain_DA} with the points used for assimilation being inside of the blue rectangles. The time domain results clearly show the improvement in the forecast accuracy of the analysis solution over the initial forecast. Using Eq.~\eqref{eq:fc_err}, we quantify the forecast accuracies in Lyapunov time units to be approximately 0.89 for the initial forecast and 4.08 for the optimized model. A learning rate decay of 0.99 was applied for obtaining these results. In subsequent sections we perform hyperparameter tuning to select optimal parameters for this system. 

\subsection{Learning Rate Dependence}\label{sec:lr_dep}
The TADA algorithm requires choosing a learning rate and learning rate decay rate for the gradient descent operation to determine the step size in the loss function space and speed up convergence of the optimization problem. For a typical application of gradient descent, learning rates are chosen near $10^{-2}$ \cite{pandey2014adaptation} because any smaller and many problems will not converge in a reasonable number of steps. However, for the TADA algorithm, the learning rates need to be chosen many orders of magnitude smaller than a traditional gradient descent problem. This is because for a chaotic system, the model coefficients are highly sensitive to changes and, if a change is significant enough, it results in a drastically different forecast prediction. Note that for all testing in this section a window size of 50 was arbitrarily chosen. So as measurements stream in, the window size grows until it reaches 50 and then slides across the signal for the final 50 steps. In total, 100 incoming measurement points are used for TADA update steps and the Adam optimizer was used for all gradient descent steps. 

\begin{figure}[H]
    \centering
    \begin{minipage}{0.45\textwidth}
        \centering
        \includegraphics[width=\textwidth]{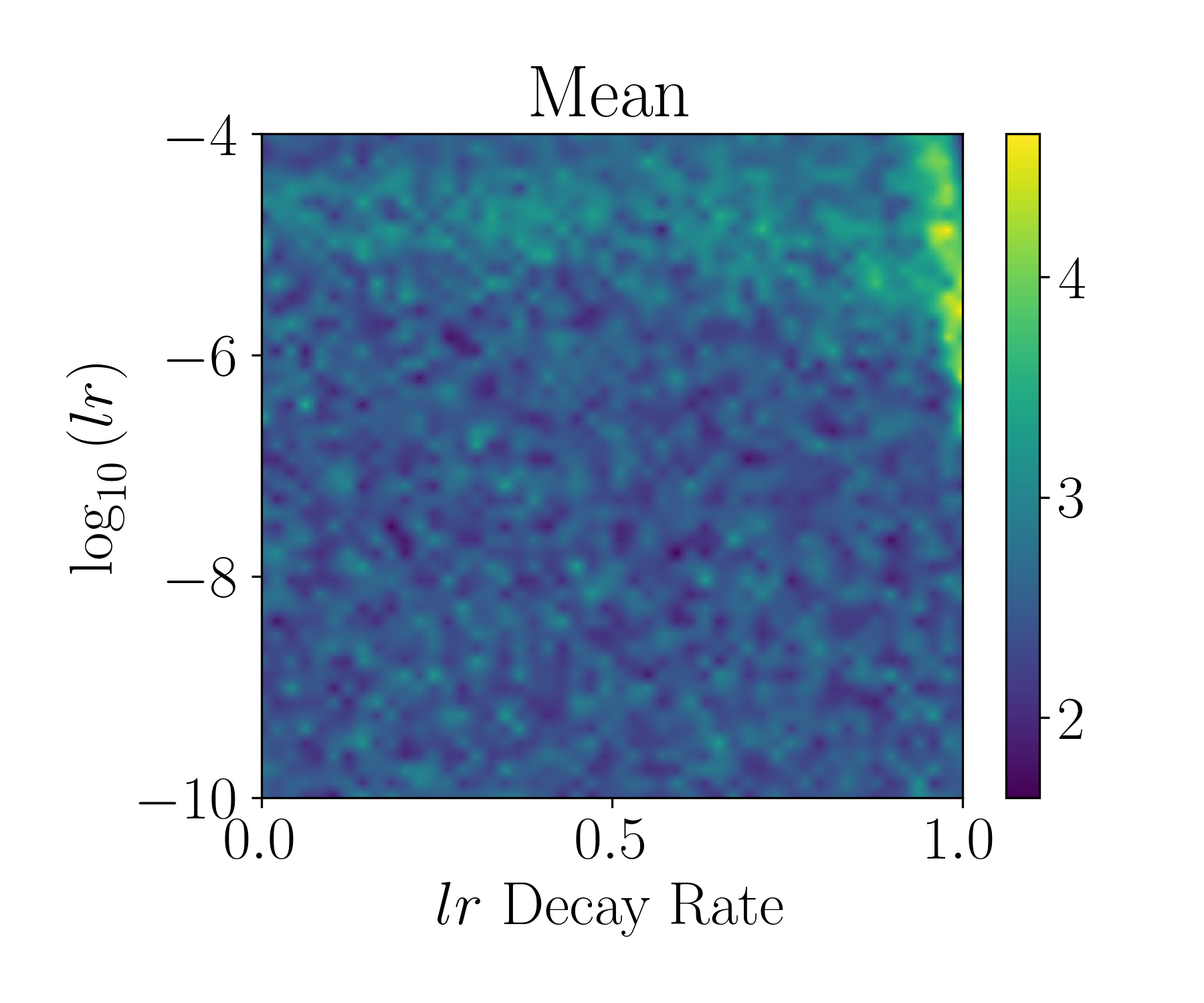}
        (a)
    \end{minipage}
    \hspace{0.05\textwidth}
    \begin{minipage}{0.45\textwidth}
        \centering
        \includegraphics[width=\textwidth]{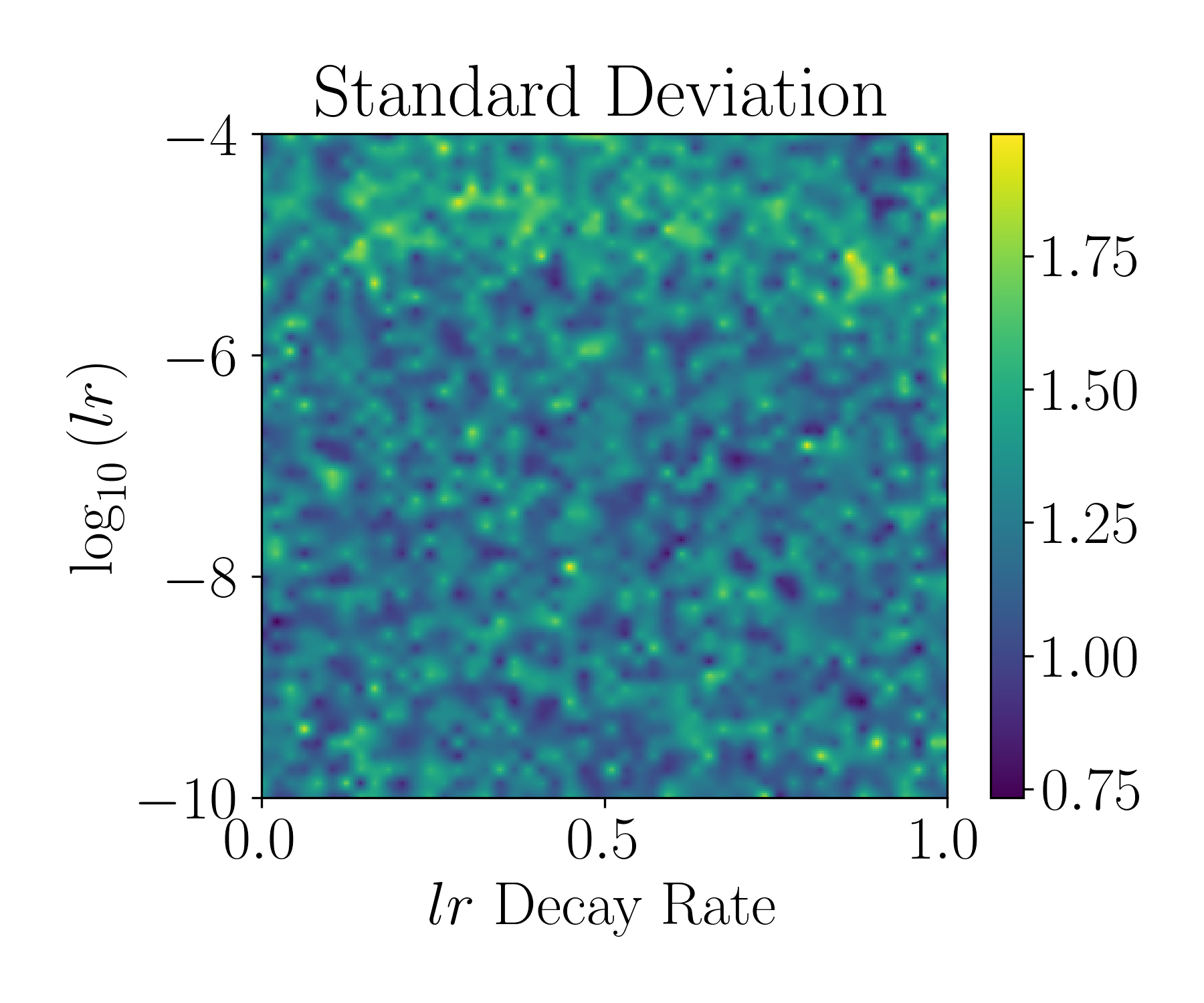}
        (b)
    \end{minipage}
    \caption{TADA ($J_1$) learning rate and learning rate decay rate analysis at for the chaotic Lorenz system. The average forecast times over 50 iterations are plotted in Lyapunov time units with respect to the TADA learning rate used on a log base 10 scale and the learning rate decay rate.}
    \label{fig:lr_lr_dec_dependence}
\end{figure}

To determine the optimal learning rate, we chose to conduct testing over a range of learning rates and learning rate decay rates at different initial conditions using noise-free signals. This test was conducted by simulating the chaotic Lorenz system at 50 different initial conditions sampled from the standard normal distribution. The mean and standard deviation forecast times were computed on $50\times 50$ grid in the learning rate and learning rate decay rate parameter space. The learning rate was varied from $10^{-10}$ to $10^{-4}$ and the decay rate was varied between 0 and 1. First, we generated these results using only the $J_1$ cost function term and the results were linearly interpolated to a resolution of $300\times 300$ and the results are shown in Fig.~\ref{fig:lr_lr_dec_dependence}. We see in Fig.~\ref{fig:lr_lr_dec_dependence}(a) that for learning rates between $10^{-6}$ and $10^{-4}$ and decay rates near one the average forecast time is more than 4 Lyapunov times whereas we will see the linear regression model forecast time is always near 2 Lyapunov times. However, the range of parameters where the forecast time increases is relatively small. 

\begin{figure}[H]
    \centering
    \begin{minipage}{0.45\textwidth}
        \centering
        \includegraphics[width=\textwidth]{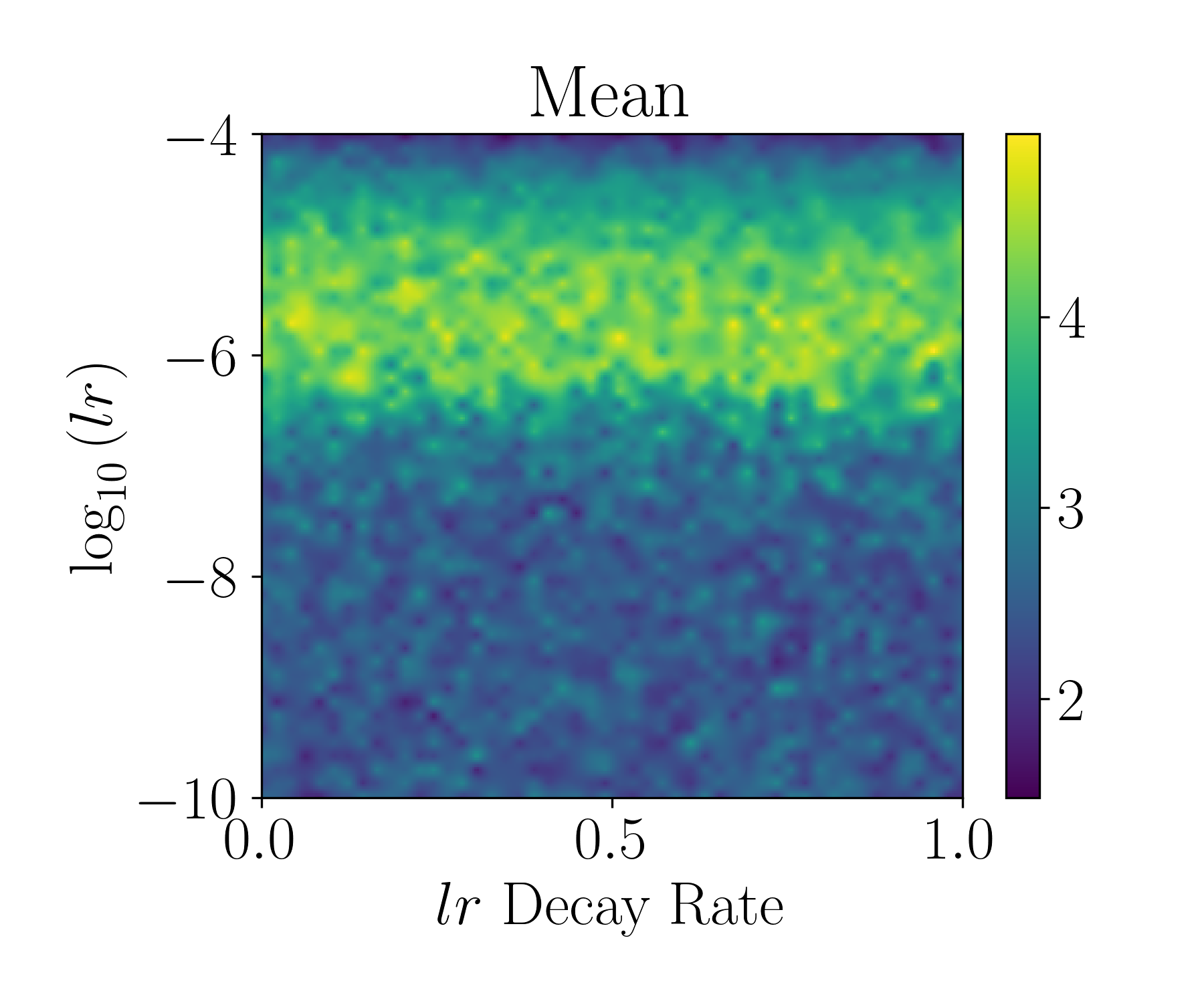}
        (a)
    \end{minipage}
    \hspace{0.05\textwidth}
    \begin{minipage}{0.45\textwidth}
        \centering
        \includegraphics[width=\textwidth]{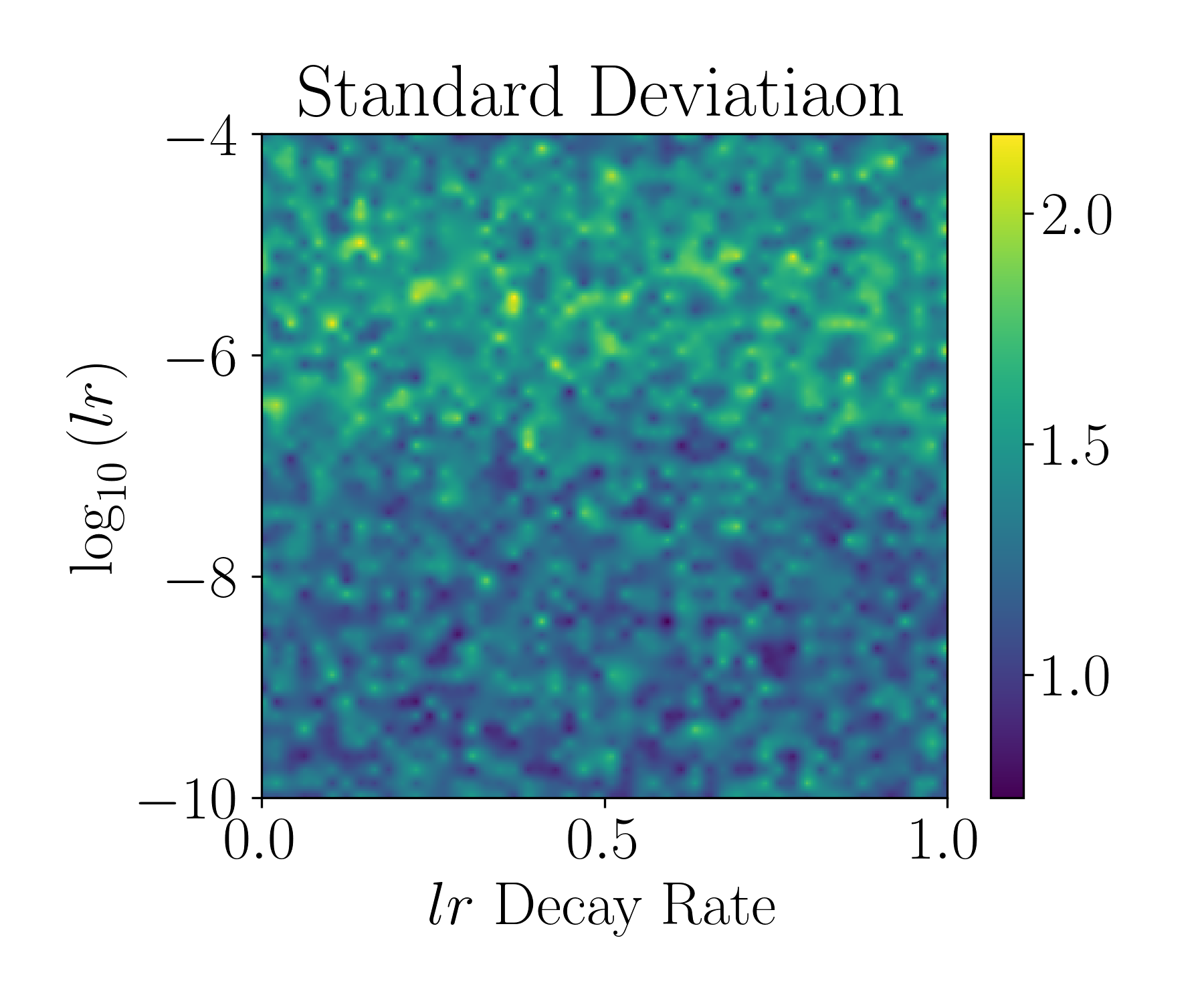}
        (b)
    \end{minipage}
    \caption{TADA ($J_1+J_2$) learning rate and learning rate decay rate analysis at for the chaotic Lorenz system. The average forecast times over 50 iterations are plotted in Lyapunov time units with respect to the TADA learning rate used on a log base 10 scale and the learning rate decay rate.}
    \label{fig:lr_lr_dec_dependence_J2}
\end{figure}

The standard deviation of the forecast times is shown in Fig.~\ref{fig:lr_lr_dec_dependence}(b) for reference. Conversely, when we introduce the $J_2$ cost function term in addition to $J_1$, the results are much more robust with larger forecast times on average as shown in 
Fig.~\ref{fig:lr_lr_dec_dependence_J2}(a) with the standard deviation in Fig.~\ref{fig:lr_lr_dec_dependence_J2}(b). We see that the average forecast time between $10^{-6}$ and $10^{-4}$ is near 4 Lyapunov times for nearly all decay rates. These results demonstrate the importance of the $J_2$ term. We chose a decay rate of 0.99 for the rest of the results.

\begin{figure}[htbp]
    \centering
    \includegraphics[width=0.5\linewidth]{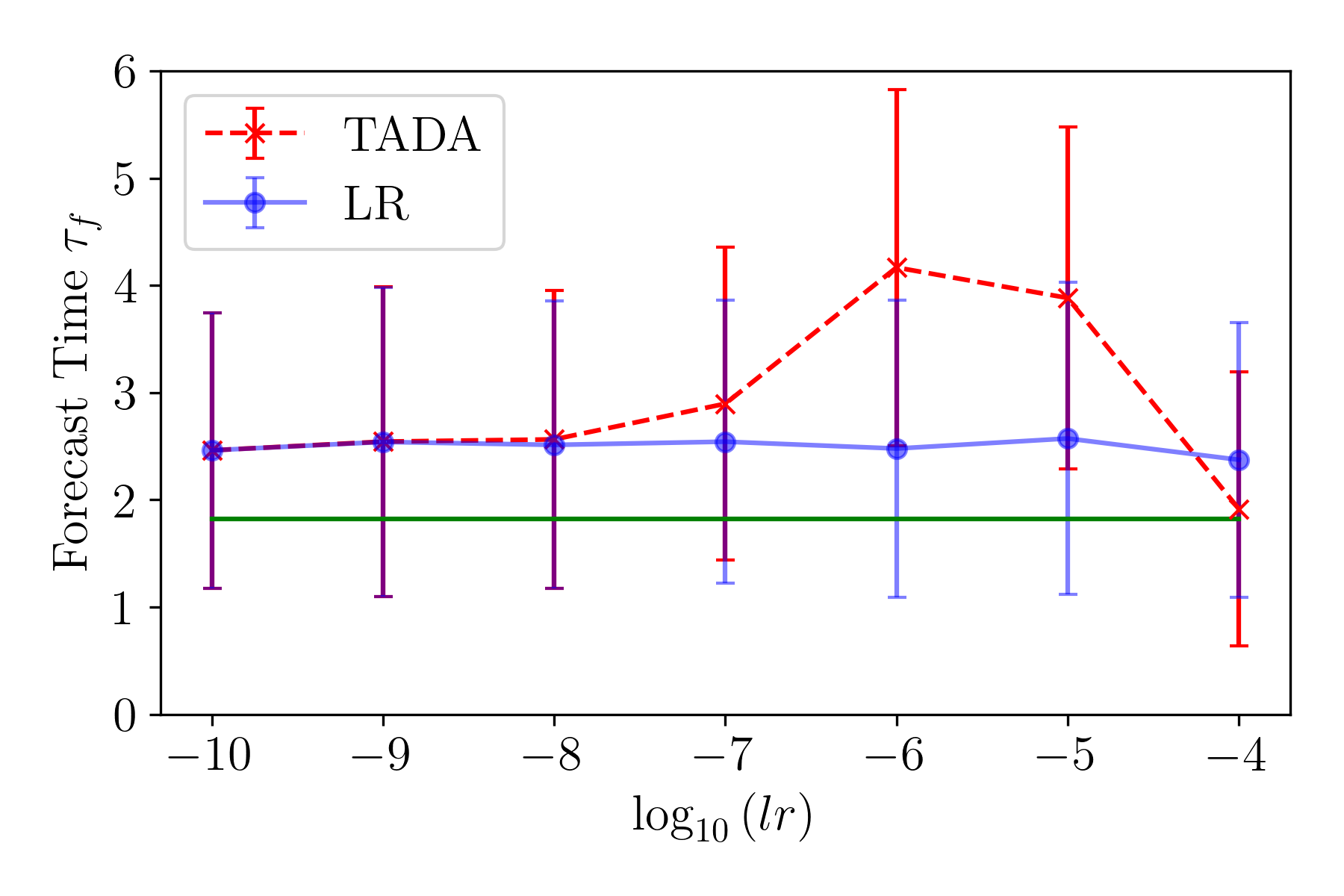}
    \caption{TADA learning rate analysis at for the chaotic Lorenz system. The average forecast times over 500 iterations are plotted in Lyapunov time units with respect to the TADA learning rate used on a log base 10 scale. Error bars indicate one standard deviation, the red x's are the TADA results and the blue points are results from random feature maps. The solid green line represents the amount of incoming data that was used to improve the model using TADA updates.}
    \label{fig:lr_dependence}
\end{figure}

Next, we compare the TADA results at varying learning rates to the random feature map Linear Regression (LR) method. This was done by testing the algorithm at a noise level of 50 dB and varying the TADA learning rate with a decay rate set to 0.99 to decrease the learning rate by 1\% for each assimilation step. Each learning rate was tested for 500 different initial conditions for this test and the results for noise free signals are shown in Fig.~\ref{fig:lr_dependence}. We see that the forecast time has the largest improvement for learning rates of $10^{-6}$ and $10^{-5}$ with $10^{-6}$ being slightly higher on average. Note the green horizontal line indicates the 100 incoming data points that were used for TADA update steps after the original training set for the LR method. We chose to explore the implications of choosing $10^{-6}$ and $10^{-5}$ for the remainder of this paper. While this method requires significant parameter tuning to get improved forecast times, this analysis can be conducted on a training set before running the TADA algorithm on a specific system and because the results for one learning rate do not depend on another it can be conducted in parallel to find the learning rate that maximizes forecast time. This process is essentially measuring the sensitivity of the model to changes in the coefficients. 

\subsection{Window Size Dependence}
\begin{figure}[htbp]
    \centering
    \includegraphics[width=0.5\linewidth]{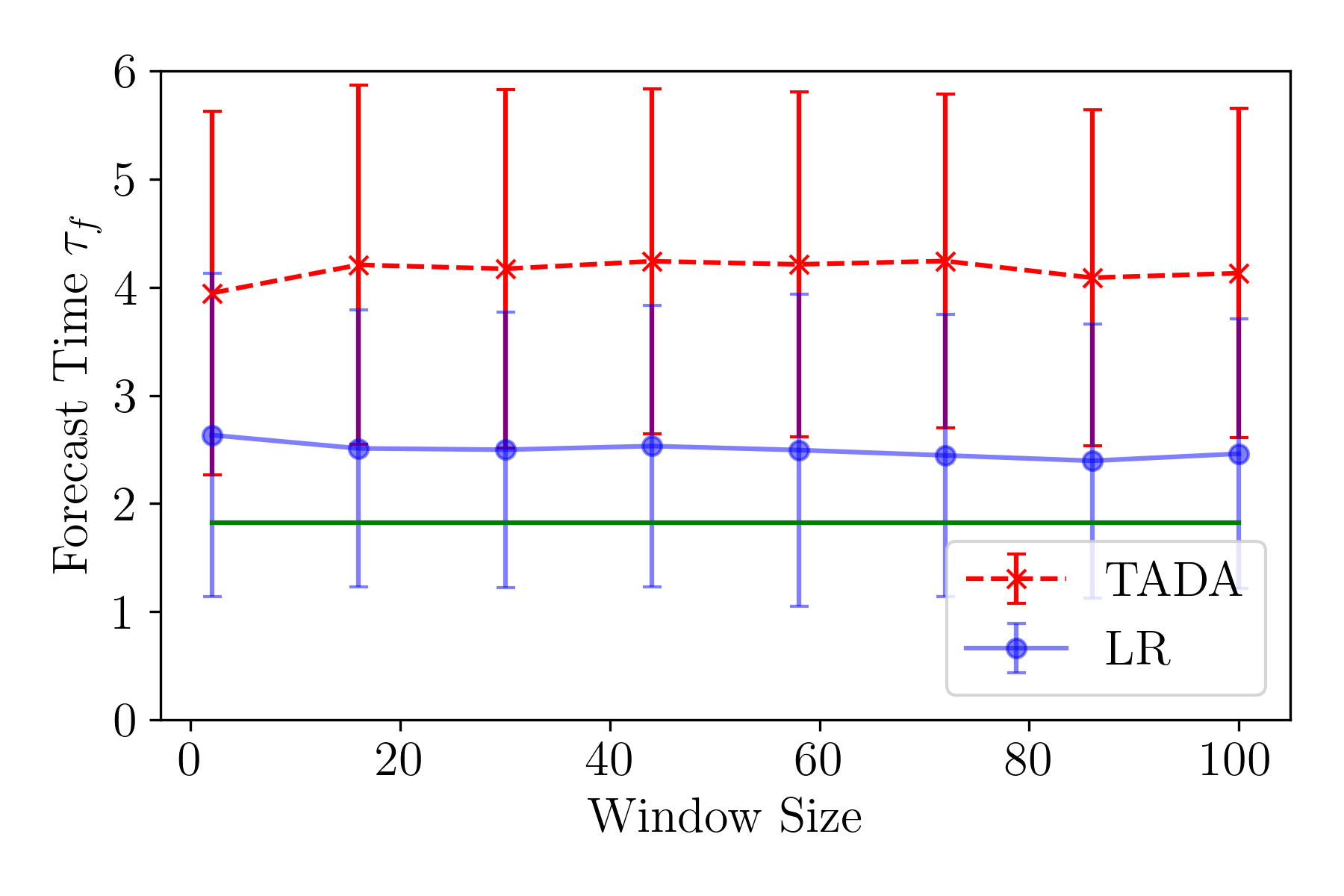}
    \caption{Average TADA forecast times plotted in Lyapunov time units with respect to the sliding window size for TADA. Red x's indicate the TADA results and blue points are the random feature map results averaged over 500 randomly chosen initial conditions with error bars indicating one standard deviation. The solid horizontal green line indicates the amount of incoming data that was used to improve the model.}
    \label{fig:window_size_dependence}
\end{figure}

To use the TADA algorithm, a sliding window width must be chosen for computing persistence. If the window size is too small, persistence features will not show up in the persistence diagrams and if the window size is too large the computation times will be too long. To determine the optimal window size, the average forecast times for the chaotic Lorenz system over 500 iterations were computed with respect to the window size and the results are shown in Fig.~\ref{fig:window_size_dependence}. A learning rate of $1\times 10^{-6}$ was used for this analysis with a decay rate of 0.99. We computed the forecast times at 500 randomly chosen initial conditions for 8 different window sizes between 0 and 100. This analysis was performed on signals with 50 dB of added white noise and the results are shown in Fig.~\ref{fig:window_size_dependence}. These results suggest that the forecast times are independent of window size as long as some persistence features are present. We found that a window size of 50 points was a good balance of computation time and significant topological features being present in the persistence diagrams, but results may vary for other systems depending on the sampling rate so this parameter can be decreased to improve computation times if needed and increased if more persistence features are required.

\subsection{Noise Robustness}\label{sec:noise_robustness}

\begin{figure}[p]
    \centering
    \begin{minipage}{0.48\textwidth}
        \centering
        \includegraphics[width=\textwidth]{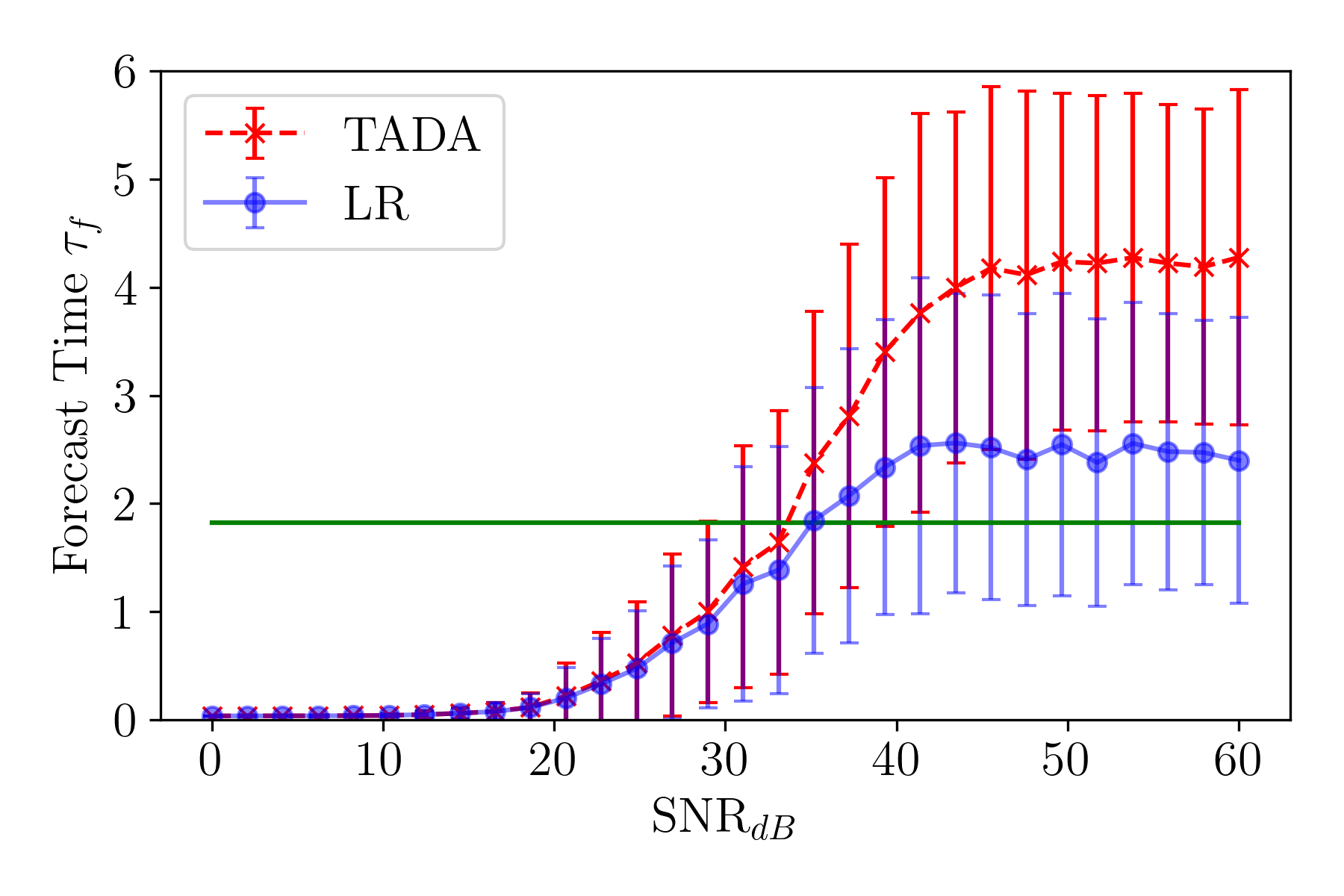}
        (a) White Noise ($10^{-6}$)
    \end{minipage}
    \begin{minipage}{0.48\textwidth}
        \centering
        \includegraphics[width=\textwidth]{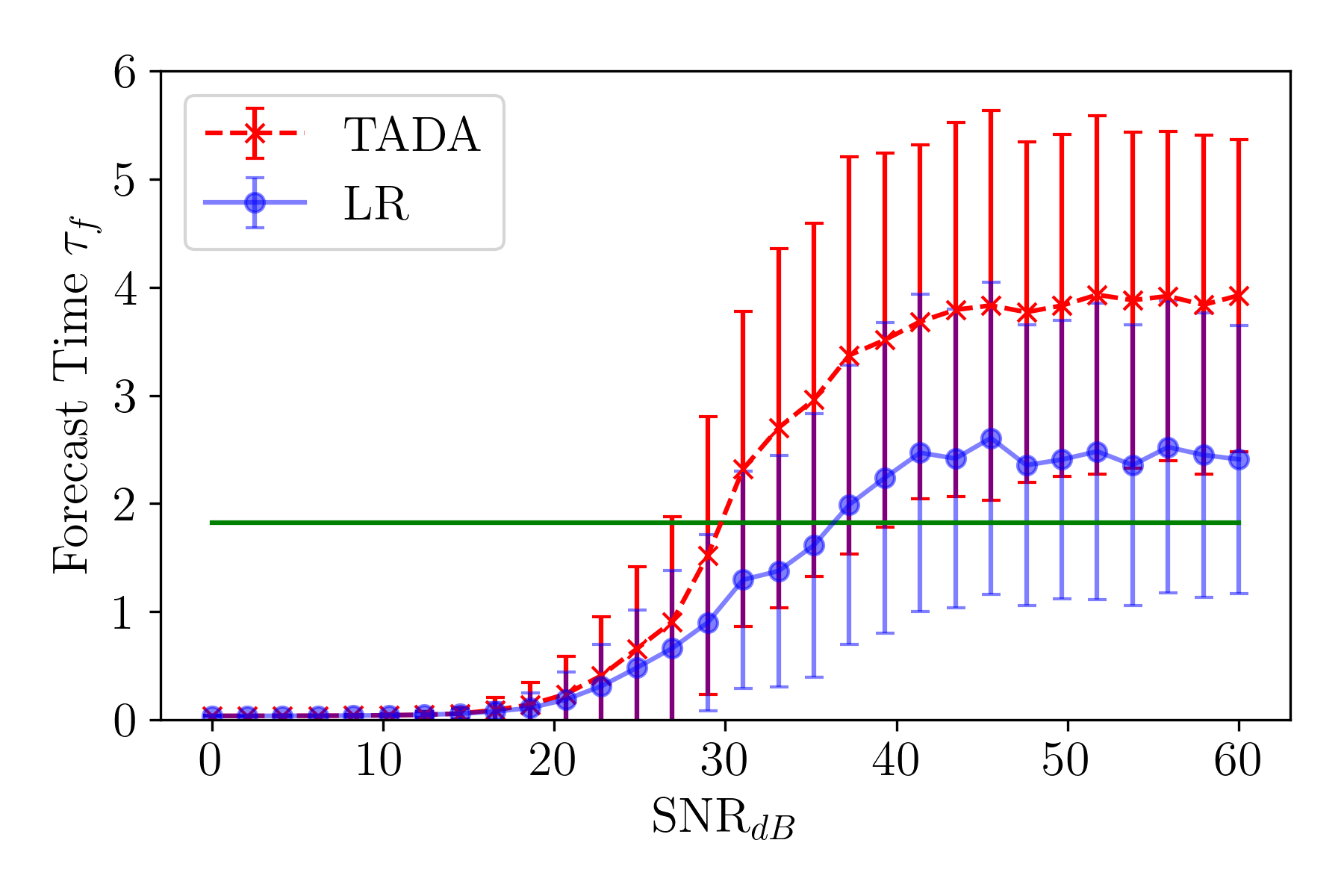}
        (b) White Noise ($10^{-5}$)
    \end{minipage}

    \begin{minipage}{0.48\textwidth}
        \centering
        \includegraphics[width=\textwidth]{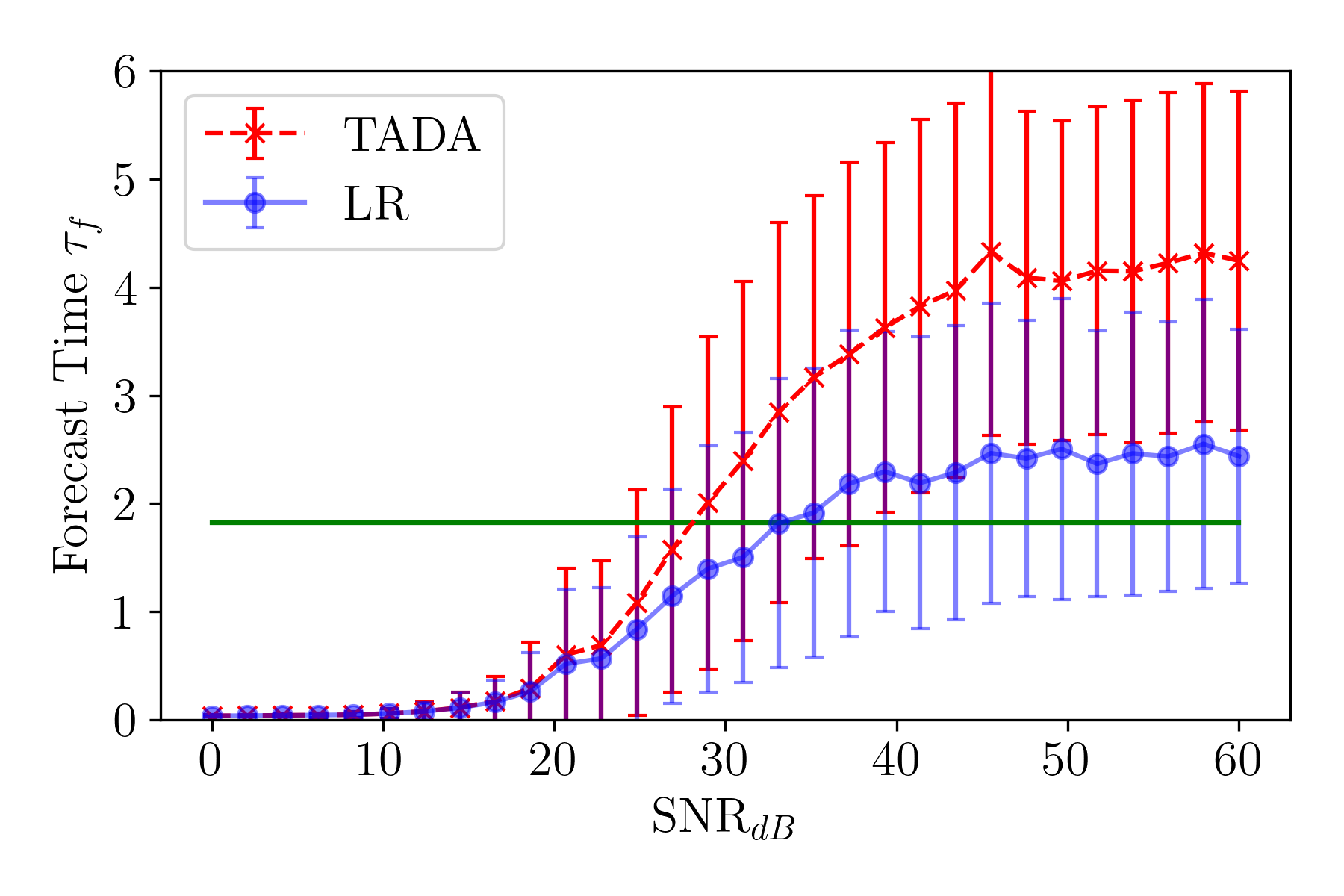}
        (c) Pink Noise ($10^{-6}$)
    \end{minipage}
    \begin{minipage}{0.48\textwidth}
        \centering
        \includegraphics[width=\textwidth]{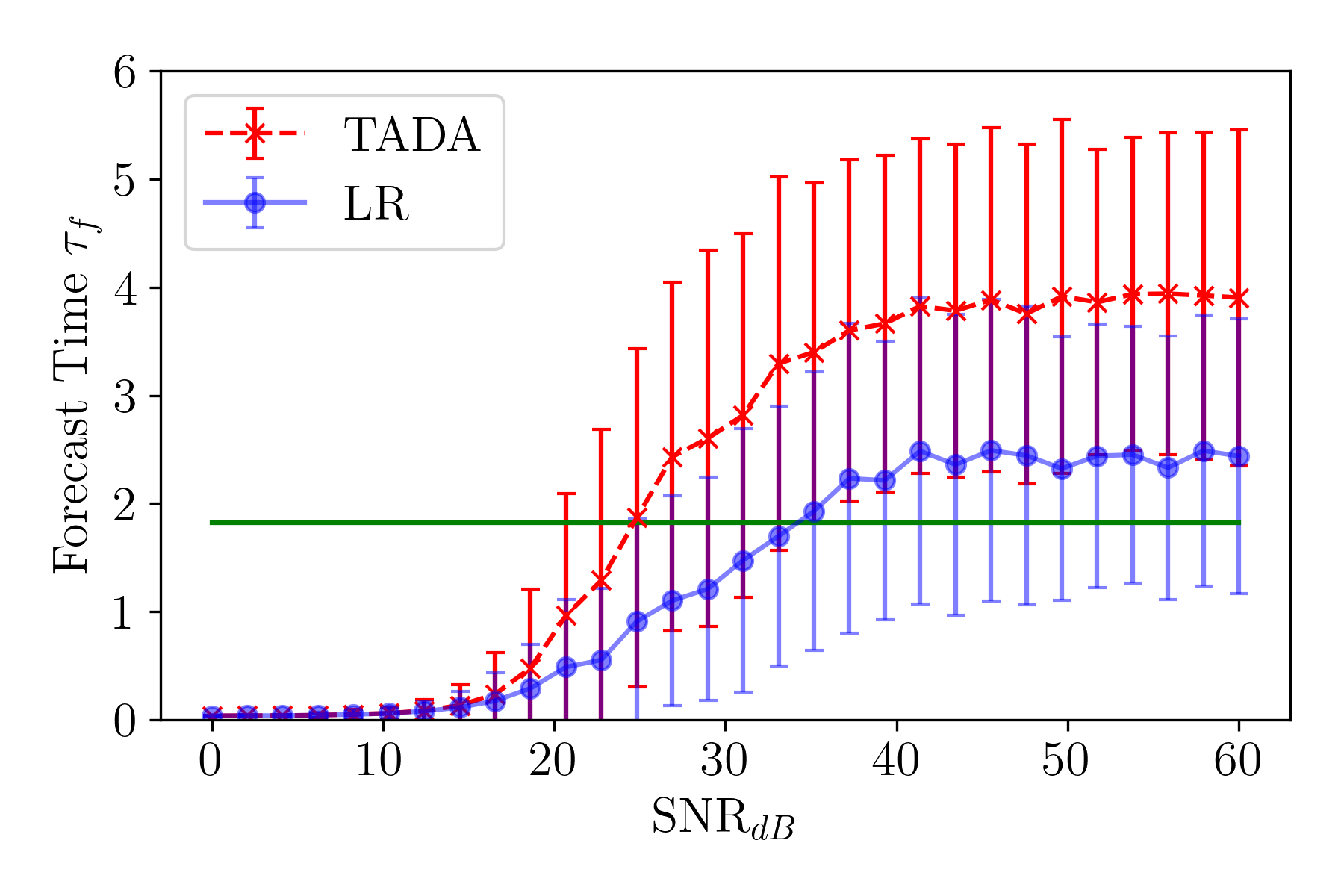}
        (d) Pink Noise ($10^{-5}$)
    \end{minipage}
    \begin{minipage}{0.48\textwidth}
        \centering
        \includegraphics[width=\textwidth]{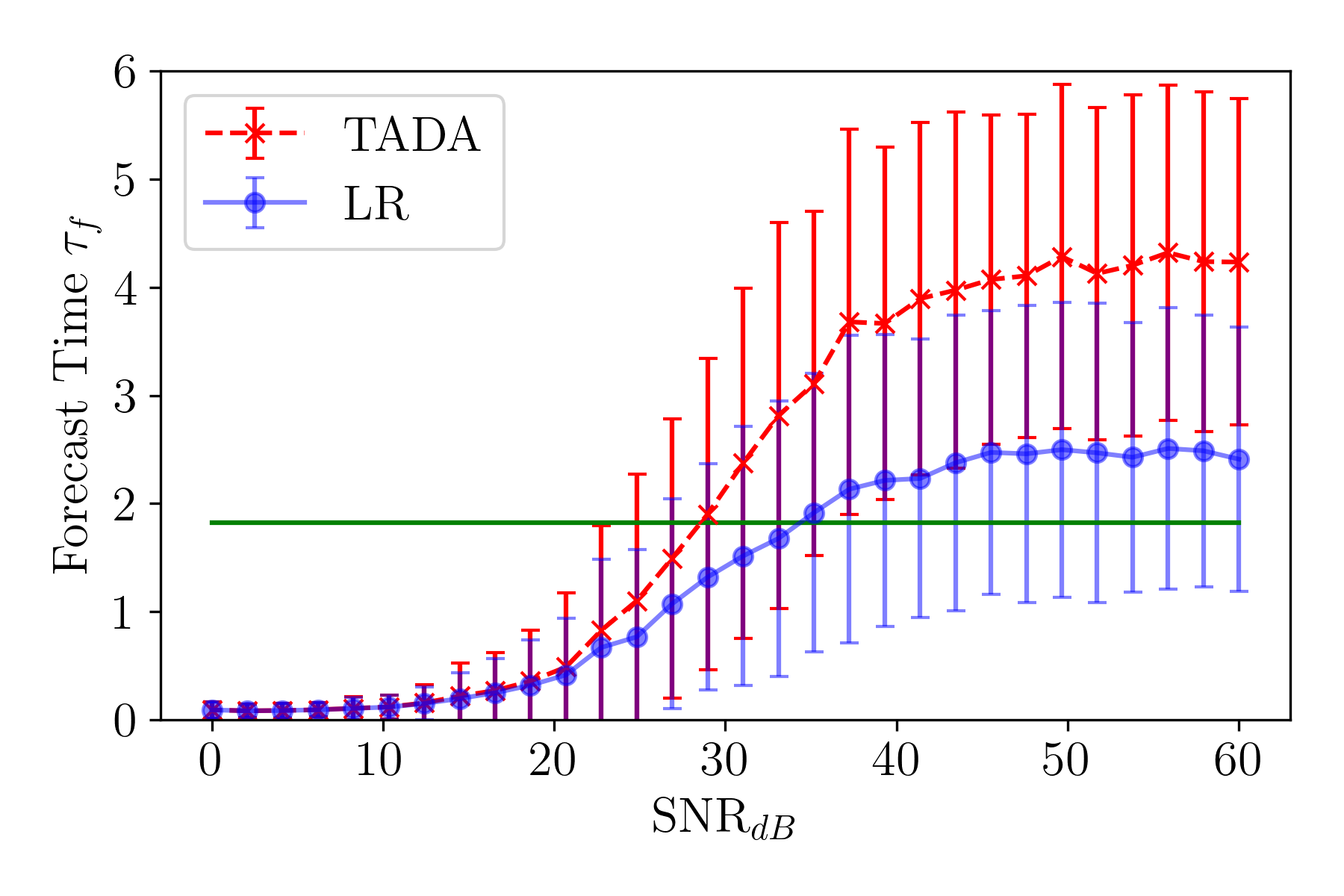}
        (e) Brownian Noise ($10^{-6}$)
    \end{minipage}
    \begin{minipage}{0.48\textwidth}
        \centering
        \includegraphics[width=\textwidth]{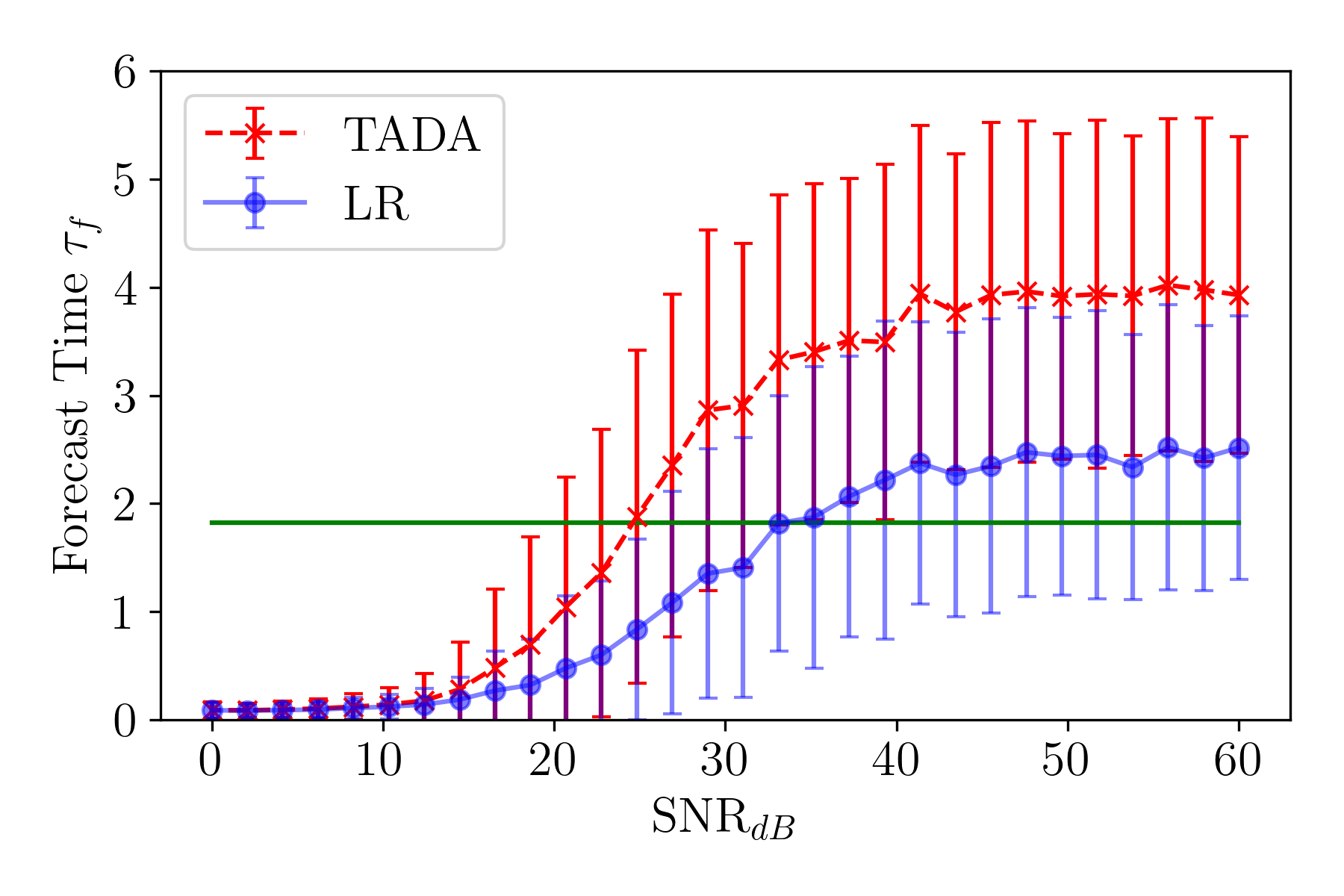}
        (f) Brownian Noise ($10^{-5}$)
    \end{minipage}

    \caption{Average TADA forecast times plotted in Lyapunov time units with respect to the SNR of the signals using three different noise distributions and two learning rates. Red x's indicate the TADA results and blue points are the random feature map results averaged over 500 randomly chosen initial conditions with error bars indicating one standard deviation. The solid horizontal green line indicates the amount of incoming data that was used to improve the model.}
    \label{fig:noise_robustness}
\end{figure}

To test the noise robustness of TADA, we measure the forecast time of the optimized models when artificial noise is added to the training signals and measurements. Many DA methods make assumptions on the noise distribution in the signal or require knowledge of the noise statistics such as in \cite{gottwald2021combining}, but TADA performs DA updates based only on topological differences between the forecast and measurements and in theory can work for any noise distribution with reasonable SNR. It is common for systems to have colored/correlated noise distributions in practice \cite{schueller2006developments, ding2013auxiliary} so the white noise assumption of other methods can be a limiting factor. 
To validate these claims, we used three different noise distributions: white, pink and brownian for our testing and applied the noise additively at varying SNR to the signals. These colored noise distributions were generated using the \texttt{colorednoise} python library which implements the algorithms from \cite{timmer1995generating}. This test was conducted over a range of 30 signal-to-noise ratios varying from 0 to 60 dB and the forecast time was measured for 500 randomly chosen initial conditions at each SNR with learning rates of $10^{-6}$ and $10^{-5}$. The results of this test are presented in Fig.~\ref{fig:noise_robustness}. For gaussian white noise, we see in Fig.~\ref{fig:noise_robustness}(a) that on average TADA improves the forecast time beyond the additional measurements used down to an SNR of approximately 32 dB with a learning rate of $10^{-6}$. However, using a learning rate of $10^{-5}$ results in slightly lower forecast times at low noise levels, but the algorithm is more noise robust down to approximately 29 dB as shown in Fig.~\ref{fig:noise_robustness}(b). Similar behavior is observed for pink and brownian noise in Fig.~\ref{fig:noise_robustness}(c-f). Interestingly when using pink and brownian noise distributions in Figs.~\ref{fig:noise_robustness} (d) and (f) TADA is more robust to noise down to an SNR of about 23 dB. Below these SNRs, the average forecast times are still improved over the LR method alone, but it does not improve the forecast time beyond the additional data that was used to improve the model (solid green horizontal line). For high SNR, the forecast time is improved on average by approximately 1.5 lyapunov times over the LR method and as more incoming measurements are used this difference increases if the correct learning rate and decay rate are chosen.

\subsection{Computation Times}

The TADA computation time was measured for the noise robustness testing in Section~\ref{sec:noise_robustness} by timing 500 iterations at each point with each iteration containing 100 DA update steps. These computations were done in parallel with 50 cores/jobs using an Intel Xeon Silver 4216 2.10 GHz CPU with 64 Gb of RAM. The average and standard deviation time for 500 iterations was computed for each noise color. To get the average iteration time we assume that each core/job performed 10 of the iterations and divide the average time by 100 steps and to get the TADA step time. In other words we divide the average time for 500 iterations by 1000 to get the average TADA step time. The resulting computation times are shown in Table~\ref{table:comp_times}. While these computation times are likely not fast enough to run the algorithm in real time, some systems may only obtain one measurement per second or even longer so in those cases this method can be used online and if the number of model parameters is optimized further the step time can be reduced.

\begin{table}[htbp]
\centering
\begin{tabular}{| c | c | c | c |}
\hline
 Noise Color & 500 Iteration Mean (sec) & 500 Iteration Standard Deviation (sec) & TADA Step Time (sec) \\
\hline
White & 1067.31 & 11.28 & \textbf{1.07}
 \\
\hline
Pink & 1069.51 & 7.11 & \textbf{1.07}
 \\
\hline
Brownian & 1077.59 & 3.55 & \textbf{1.08}
 \\
\hline
\end{tabular}
\caption{TADA Computation Times over 500 iterations for each noise distribution tested. The DA step time was determined by assuming that each of the 50 cores used in the parallel computation handled 10 of the 500 iterations and that time was then divided by the 100 DA steps in each iteration.}
\label{table:comp_times}
\end{table}

\subsection{Lorenz 96 Example}

\begin{figure}[t]
    \centering
    \begin{minipage}{0.49\textwidth}
        \centering
        \includegraphics[width=\textwidth]{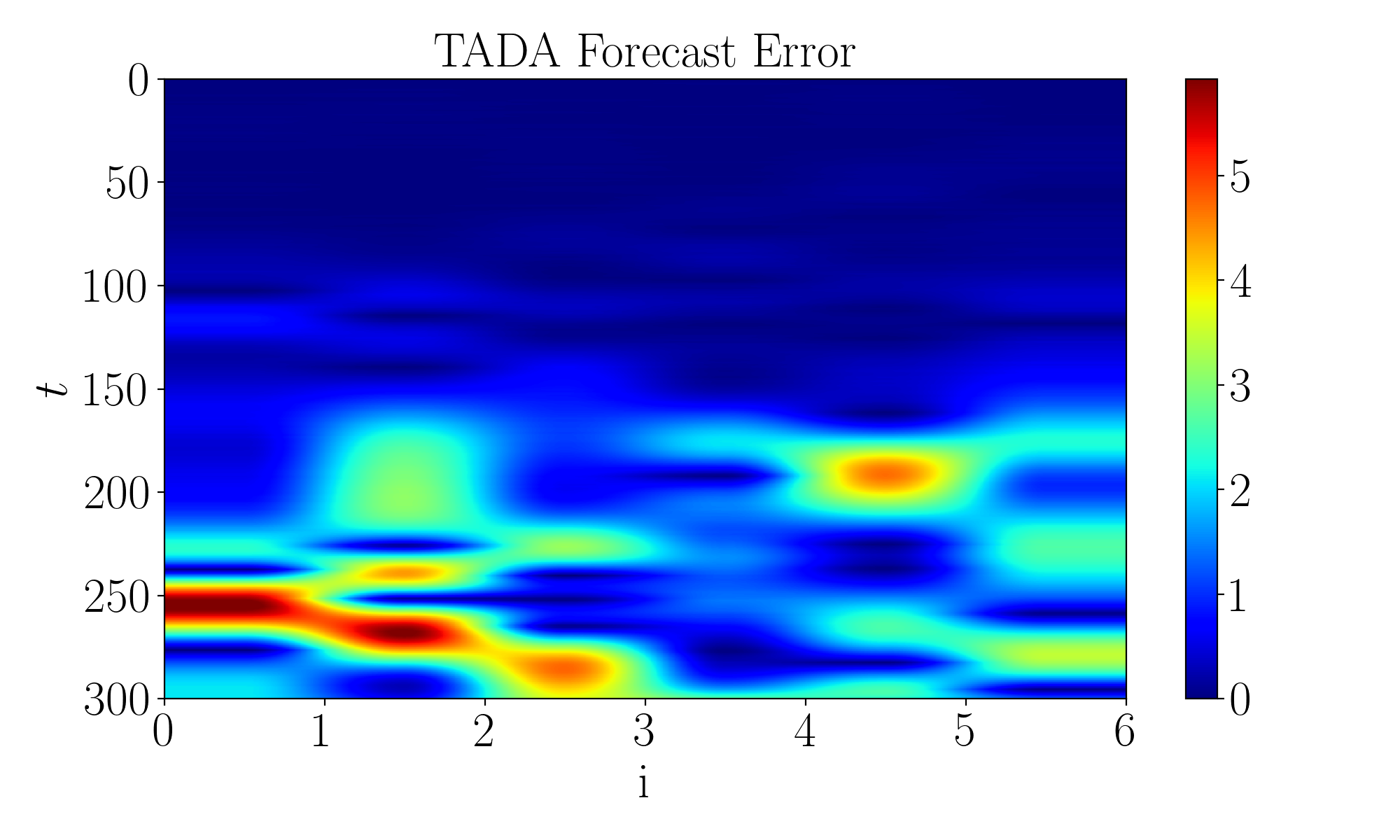}
        (a) Initial Error
    \end{minipage}
    \begin{minipage}{0.49\textwidth}
        \centering
        \includegraphics[width=\textwidth]{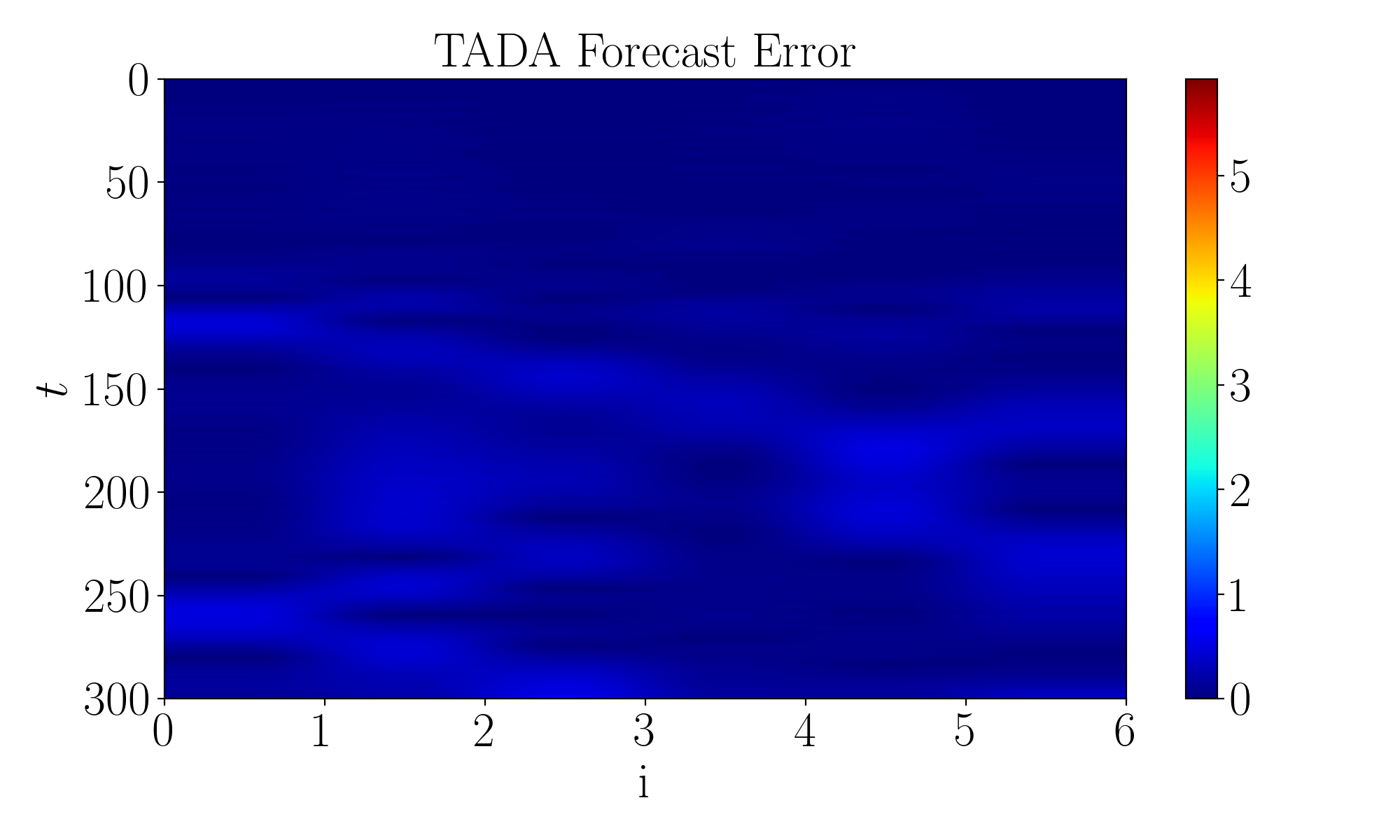}
        (b) Final Error
    \end{minipage}
    \caption{(a) Initial and (b) final 6 dimensional Lorenz 96 TADA error between the updated TADA model and the measurements after 100 TADA update steps at a learning rate of $1\times 10^{-6}$.}
    \label{fig:L96_6dim_err}
\end{figure}

All of the results to this point have been from the 3D Lorenz 63 system. Here we demonstrate the TADA forecast capability on a higher dimensional system by generating optimal models for a variant of the Lorenz 96 system. The Lorenz 96 model is given by:
\begin{equation}
    \frac{dX_i}{dt} = (X_{i+1} - X_{i-2}) X_{i-1} - X_i + F,
\end{equation}
where $X_i$ is the $i$-th state, with $i=1,\cdots,N$, $F$ is the forcing parameter and $X_{-1}=X_{N-1}$, $X_0=X_N$, $X_{N+1}=X_1$ with $N\geq 4$ \cite{karimi2010extensive}. Interestingly, this system allows for specifying the dimension as a system parameter. Physically, this system represents climate dynamics on a circle and we use $F=8$ to yield a chaotic response \cite{ott2004local}. For these results, we choose $N=6$ to start and rather than using the forecast time from Eq.~\eqref{eq:fc_err} we use the absolute difference between the states and measurements to plot the error in space and time. We simulated the Lorenz 96 system for 200 time units using a time step of 0.01 and removed 25 time units to allow transient behavior to dissipate. 170 time units were used for training the random feature map model with $w=0.01$, $b=0.1$ and a reservoir dimension of $D_r=500$ was used. White noise was added to the signals with an amplitude of $\eta=0.01$ and 100 TADA steps were taken at a learning rate of $1\times 10^{-6}$ with a decay rate of 0.99. The initial and final forecast absolute difference errors are shown in Fig~\ref{fig:L96_6dim_err} (a) and (b) respectively. We see that the initial forecast had low error out to about 150 time steps and after 100 TADA steps the model and measurement discrepancy approached zero out to 300 time steps. We note that this example is presented to demonstrate that our method works on higher dimensional systems, but we emphasize the importance of the hyperparameter tuning for this method as arbitrarily setting the parameters will likely yield poor results. 

%% file: Sections/conclusion.tex
\section{Conclusions}\label{sec:conclusion}
This paper initiates a numerical study for a Topological Approach for Data Assimilation (TADA). Specifically we leverage persistent homology and its nascent differentiability framework for optimally combining model predictions and measurements to update a dynamical system's model and improve prediction times. The algorithm starts with an initial data driven model and as new measurements stream in the topological differences between the forecast and measurements are minimized using persistence differentiation and gradient descent. TADA does not make any assumptions on the noise distribution of the measurements and uses a sliding window approach to reduce computation times. Numerous parameter studies were performed to validate the TADA algorithm and we demonstrate its robustness to pink, brown, and white noise terms. All testing was performed on models obtained using the random feature map method, but in the future we plan to generalize the approach to other forecast functions and test the algorithm with respect to other models such as Long Short-Term Memory (LSTM) networks.